\documentclass[twocolumn,showpacs,preprintnumbers,amsmath,amssymb,pra,superscriptaddress]{revtex4}
\usepackage{graphicx}
\usepackage{dcolumn}
\usepackage[tight]{subfigure}
\usepackage{amsmath}
\usepackage{verbatim}
\usepackage[usenames,dvipsnames]{color}
\usepackage{bm} 
\usepackage{bbm}
\usepackage{floatrow}

\newcommand{\Figure}[2]{
  \begin{figure}[t]
    \includegraphics[width=1.0\linewidth]{#1}
    \caption{#2}
  \end{figure}
}

\newcommand{\FigureBottom}[2]{
  \begin{figure}[b]
    \includegraphics[width=1.0\linewidth]{#1}
    \caption{#2}
  \end{figure}
}

\newcommand{\FigureSmaller}[2]{
  \begin{figure}[t]
    \includegraphics[width=0.7\linewidth]{#1}
    \caption{#2}
  \end{figure}
}

\newcommand{\Bigfigure}[2]{
  \begin{figure*}[ht]
    \includegraphics[width=1.0\linewidth]{#1}
    \caption{#2}
  \end{figure*}
}

\newcommand{\wideeq}[1]{
\begin{widetext}
    #1
\end{widetext}}

\newcommand{\Fig}[1]{Fig.~\ref{#1}}

\newcommand{\eq}[1]{(\ref{#1})}
\newcommand{\Sec}[1]{Sec.~\ref{#1}}
\renewcommand{\sec}[1]{\ref{#1}}
\newcommand{\App}[1]{App.~\ref{#1}}

\newcommand{\Cite}[1]{Ref.~\onlinecite{#1}}

\newcommand{\cut}[1]{\textcolor{red}{cut: #1}}
\newcommand{\todo}[1]{ \textbf{\textcolor{Bittersweet}{TODO: #1}} }
\newcommand{\hide}[1]{ \textbf{\textcolor{Gray}{#1}} }

\newcommand{\tmem}[1]{{\em #1\/}}
\newcommand{\tmop}[1]{\ensuremath{\operatorname{#1}}}

\newcommand{\nobracket}{}

\renewcommand{\cut}[1]{{}}
\renewcommand{\todo}[1]{}
\renewcommand{\hide}[1]{}

\renewcommand{\vec}[1]{\mathbf{#1}}

\newcommand{\tmmathbf}[1]{\ensuremath{\boldsymbol{#1}}}

\begin{document}
\title{Coupling and braiding Majorana bound states\\ in networks defined in proximitized two-dimensional electron gases}
\author{Michael Hell}
\affiliation{Division of Solid State Physics and NanoLund, Lund University, Box.~118, S-22100, Lund, Sweden}
\affiliation{Center for Quantum Devices and Station Q Copenhagen, Niels Bohr Institute, University of Copenhagen, DK-2100 Copenhagen, Denmark}
\author{Karsten Flensberg}
\affiliation{Center for Quantum Devices and Station Q Copenhagen, Niels Bohr Institute, University of Copenhagen, DK-2100 Copenhagen, Denmark}
\author{Martin Leijnse}
\affiliation{Division of Solid State Physics and NanoLund, Lund University, Box.~118, S-22100, Lund, Sweden}
\affiliation{Center for Quantum Devices and Station Q Copenhagen, Niels Bohr Institute, University of Copenhagen, DK-2100 Copenhagen, Denmark}
\date{\today}

\begin{abstract}
  Two-dimensional electron gases with strong spin-orbit coupling covered by a
  superconducting layer offer a flexible and potentially scalable platform for
  Majorana networks. We predict Majorana bound states (MBSs) to appear for
  experimentally achievable parameters and realistic gate potentials in two designs: either
  underneath a narrow stripe of a superconducting layer (S-stripes) or where a
  narrow stripe has been removed from a uniform layer (N-stripes). The
  coupling of the MBSs can be tuned for both types in a wide range ($< 1~$neV to
  $>10~\mu$eV) using gates placed adjacent to the stripes. For both
  types, we numerically compute the local density of states for two parallel
  Majorana-stripe ends as well as Majorana trijunctions formed in a
  tuning-fork geometry. The MBS coupling between parallel Majorana stripes
  can be suppressed below 1 neV for potential barriers in the meV
  range for separations of about 200 nm. We further show that the MBS
  couplings in a trijunction can be gate-controlled in a range similar to the
  intra-stripe coupling while maintaining a sizable gap to the excited states
  (tens of $\mu$eV). Altogether, this suggests that braiding can carried out
  on a time scale of 10-100 ns.
\end{abstract}

 \maketitle

Majorana bound states (MBSs) are quasiparticles in superconductors that are
their own 'self-adjoints'
{\cite{AliceaReview,FlensbergReview,BeenakkerReview,TewariReview}}. This
requires them to be an equal superposition of particles and holes and ties
their energy to the middle of the superconducting gap. MBSs
can appear in spatially separate pairs as, for example, at the opposite edges of a topological superconductor.
This nonlocality may be utilized for storage and manipulation of quantum
information in a topologically protected way
{\cite{BravyiKitaev,BravyiKitaev2,Freedman03}}. However, the realization of
MBSs requires superconducting p-wave pairing, which appears intrinsically only
in exotic materials. Fortunately, p-wave pairing can also be engineered by
combining s-wave superconductors with strong spin-orbit materials
{\cite{1DwiresOreg,Sau,1DwiresLutchyn,Alicea}}. Based on this, experiments
looked so far for evidence of MBSs in, for example, semiconducting nanowires
{\cite{mourik12,das12,finck12,Rokhinson,deng12,Albrecht2016,Zhang16,Deng16}},
topological insulators {\cite{Williams}}, magnetic atom chains
{\cite{Nadj-Perge,Ruby15}}, and recently also two-dimensional electron gases
{\cite{Suominen17}}.

This progress motivates further experiments that would be more conclusive than
the 'local' Majorana features seen in tunneling spectroscopy so far.
Theoretical proposals for probing their nonlocal properties range from
interference experiments
{\cite{Fu4Pi,Benjamin10,Sun14,Ueda14,Yamakage14,Sau15Nonlocal,Rubbert16,Tripathi16}},
teleportation {\cite{MajoranaTransportWithInteractions1}}, fusion-rule tests
{\cite{Aasen16}}, coherence measurement of topological qubits
{\cite{Aasen16}}, and ultimately to braiding
{\cite{AliceaBraiding,ClarkeBraiding,HalperinBraiding,BeenakkerBraiding,Aasen16,Hell16,SauBraiding,BraidingWithoutTransport,BondersonBraiding,Vijay16,Karzig16}}.
The latter would unambiguously demonstrate non-Abelian exchange statistics.
Realizing these proposals calls for a flexible platform for building complex
and controllable Majorana devices. Such a Majorana platform should preferably
also be scalable to build large-scale MBS networks later on as a central part
of a topological quantum computer.

A potential platform granting such flexibility and sca-

\Figure{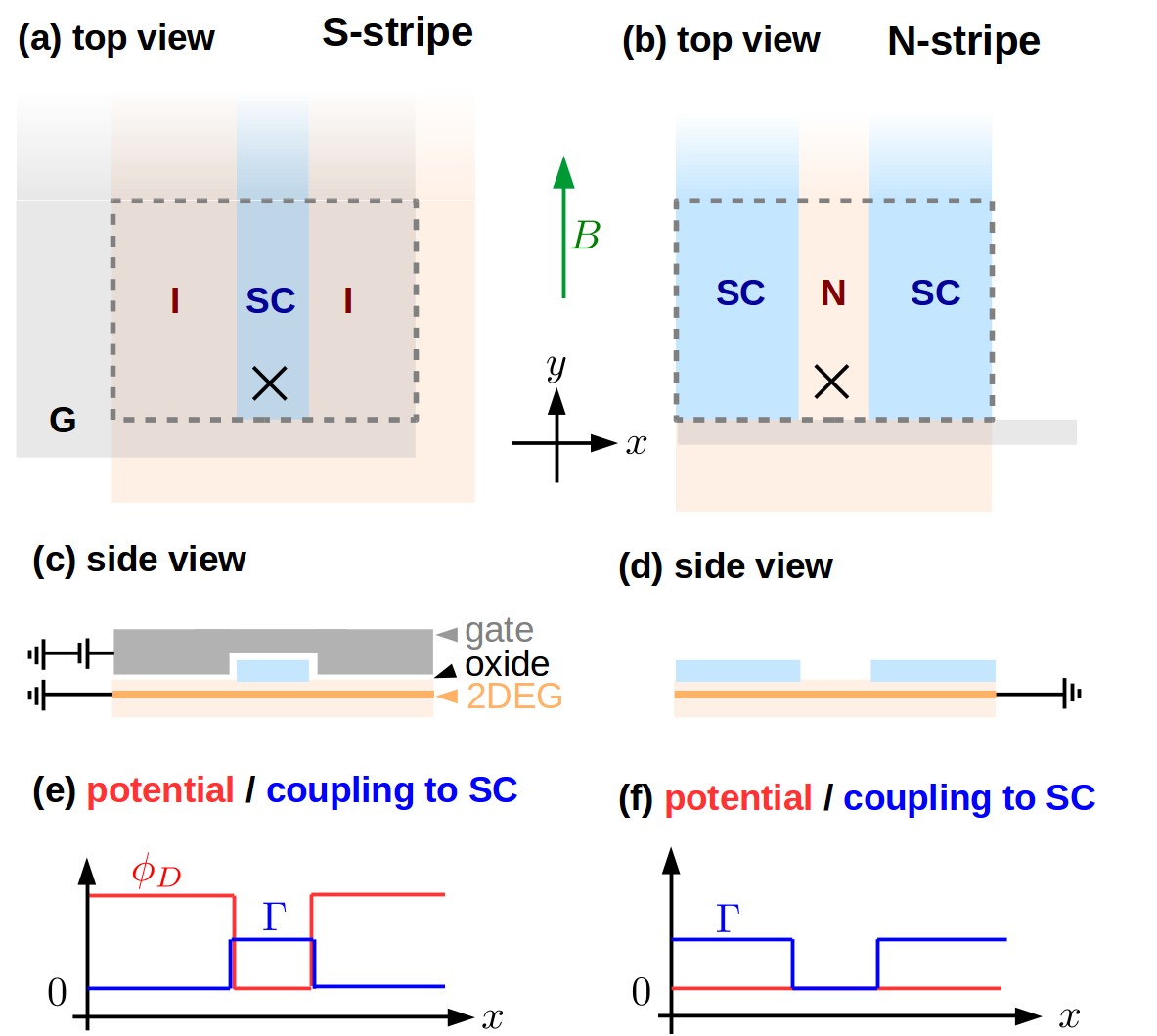}{Two
    routes to realize topological superconducting channels in
    2DEG-superconductor heterostructures. S-stripe (left panels): MBSs are
    formed underneath a stripe of a superconductor (SC) on top of the 2DEG. A
    depletion gate (G) turns the 2DEG outside the stripe insulating (I) and
    confines the states in transverse direction. N-stripe (right panels): MBSs
    are confined in a normal conducting 2DEG stripe (N)
    sandwiched between two
    2DEG regions with proximity-induced superconductivity. For both types,
    MBSs occur at the ends of the stripes as indicated by crosses (the other
    end of the stripe is not shown here). Panels (a)
    and (b) show device sketches from above and (c) and (d) from the side. The
    electrostatic confinement potential $\phi ( x)$ for the electrons (red) is shown
    in (e) and (f) alongside the tunnel coupling $\Gamma ( x)$ to the
    superconducting top layer (blue).\label{fig:devices-simple}}\clearpage
\noindent lability is based on a two-dimensional electron gas (2DEG) with strong spin-orbit coupling
{\cite{Shabani16,Suominen17}}. 
With a superconducting layer on top, the 2DEG
acquires a superconducting proximity effect, which has been investigated in
depth in the past
{\cite{SchaepersBook,Chrestin97,Chrestin99,Takayanagi95,Bauch05}}. Crucial
steps towards topological superconductivity are mainly due to advances in
material growth: By growing an Al top layer epitaxially
{\cite{Shabani16}}, clean interfaces between the 2DEG and the superconductor
can be formed so that the 2DEG develops a {\tmem{hard}} superconducting gap
{\cite{Kjaergaard16,Suominen16,KjaergaardMAR}}. Recent experiments indicate
the presence of MBSs in these heterostructures through a stable zero-bias
conductance peak {\cite{Suominen17}}. Owing to well-developed top-down
fabrication techniques, 2DEG-based Majorana devices are rather easy to
fabricate as compared to nanowire-based devices. Hence, more complex device
structures such as two-path interferometers, junctions, or trijunctions come
within reach.

To realize MBSs in such 2DEG structures, one can pursue the two routes
sketched in \Fig{fig:devices-simple}: The first approach is to fabricate
a thin stripe of aluminum with a top gate that depletes the normal conducting
2DEG around the stripe, turning it into an insulator
[\Fig{fig:devices-simple}(a)]. In this way, a narrow, quasi-1D superconducting
channel is formed under the aluminum (called S-stripe in the following). This
corresponds to the device design of the recent experiment {\cite{Suominen17}}.
The other, complementary approach {\cite{Hell17a,Pientka16}} uses a normal
conducting channel (called N-stripe in the following) next to two proximitized
superconducting regions (SC) [\Fig{fig:devices-simple}(b)].

Out of these promising developments the question arises which requirements
have to be satisfied when designing more complex Majorana devices in 2DEGs. In
this paper, we therefore numerically analyze two simple device elements both
for N- and S-stripes: two parallel Majorana stripes [Figs.~\ref{fig:devices-simulated}(a) and (b)] and a trijunction in a tuning-fork
design [Figs.~\ref{fig:devices-simulated}(c) and (d)]. The latter design
respects the key requirement that all Majorana stripes have to be parallel to
stay in the topological regime {\cite{Hell17a}}. We employ a Green's
function approach to compute the local density of states for these devices
numerically as explained in \Sec{sec:model}. This provides relevant
information about the excitation spectrum.

Using material parameters taken from experiments, we first investigate in
{\Sec{sec:stripe-single}} single N- and S-type stripes. In both cases,
we predict that the channels develop a zero-energy mode when a magnetic field
is applied along the stripe. This is a signature of a topological phase transition and formation of a MBS, similar to nanowire setups. Since single N-stripes have recently been
investigated theoretically elsewhere {\cite{Pientka16,Hell17a}}, we
focus our attention mostly to S-stripes. We verify that the effect of
realistically smoothened gating potentials hardly affects the energy spectrum,
provided the chemical potential can be tuned in the 2DEG. We further show that
the potential barrier height affects the topological phase-transition point in
S-stripes. This can be exploited to tune the MBS coupling in S-stripes
electrically, similar to a recent proposal for N-stripes {\cite{Hell17a}}.

We then show in Sec.~{\sec{sec:stripe-parallel}} that the MBSs in two parallel
stripes can be well separated from each other by applying potential
differences of a few meV for stripe distances of 200 nm or more. We further
show in \Sec{sec:tuning-fork} how to electrically control MBSs in a
trijunction region. A key question is here whether the coupling energies can
be tuned sufficiently without introducing unwanted, low-energy excitations. We
show, for a specific geometry, that the coupling energies can be tuned
between $< 1~$neV and about $10~\mu$eV, while excited states remain at tens of
$\mu$eV throughout the entire tuning range.

\Figure{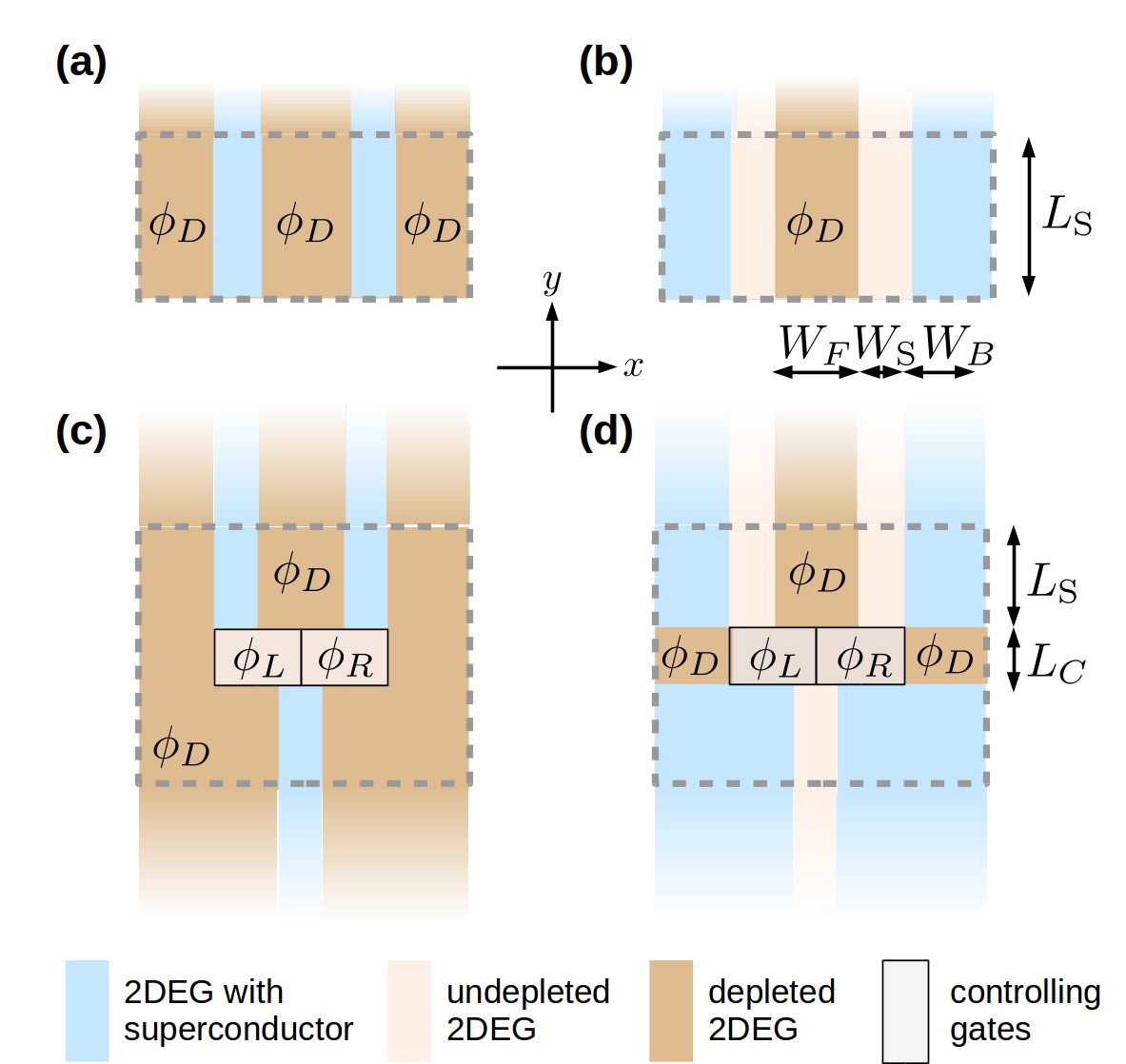}{Basic
    elements of more complex Majorana devices in 2DEGs: (a) two parallel
    S-stripe ends, (b) two parallel N-stripe ends, (c) trijunction of
    S-stripes, and (d) trijunction of N-stripes. All devices have a finite
    extension in $x$-direction transverse to the stripes, $| x | < W_F / 2 +
    W_S + W_B$, and where fainted extensions are shown, the stripes extend
    infinitely along the $y$-direction.\label{fig:devices-simulated}}
We finally address in \Sec{sec:braiding} the question how to braid MBSs
in a 2DEG-based platform both for N- and S-stripes. There are numerous
proposals for MBS braiding, which fall into three categories: The first
suggestions aimed at {\tmem{moving}} topological phase boundaries
{\cite{AliceaBraiding,ClarkeBraiding}} {\tmem{physically}}, which, however,
produces quasiparticles that prevent quantum-error correction
{\cite{PedrocchiNoise1,PedrocchiNoise2}}. In a second class of approaches, the
topological phase boundaries are preserved and braiding is achieved by
{\tmem{adiabatic manipulation}} of the couplings between the MBSs
{\cite{HalperinBraiding,BeenakkerBraiding,Aasen16,Hell16,SauBraiding,BraidingWithoutTransport}}.
Finally, the same operation as that of braiding can also be achieved by
sequences of {\tmem{measurements}} of different MBS pairs
{\cite{BondersonBraiding,Vijay16,Karzig16}}. In this paper, we mostly focus on an adiabatic braiding approach closely
related to Refs. {\cite{Aasen16}} and {\cite{SauBraiding}}. From our numerical
results for the excitation spectrum, we estimate a time scale for braiding of about 10-100 ns.

\section{Model and numerical approach}\label{sec:model}

To model the six devices shown in Figs.~\ref{fig:devices-simple} and
\ref{fig:devices-simulated}, we use a Green's function formalism. In \Sec{sec:ham}, we first introduce the 2DEG Green's function, containing the
2DEG Hamiltonian and the self-energy contribution from the superconducting top
layer, which has been integrated out. In contrast to many other studies, we
keep finite-frequency corrections to obtain a more accurate description of the
energy spectrum. For our numerical calculations, we introduce a lattice and
compute the local density of states in the finite regions framed by dashed
lines in Figs.~\ref{fig:devices-simple} and \ref{fig:devices-simulated}. The
fainted parts outside this region are assumed to continue infinitely in
vertical direction. These are accounted for by a surface Green's function,
which is nonzero on the boundary of the framed regions. Details of our
algorithm are relegated to \App{app:gf}.

\subsection{2DEG with proximity-induced superconductivity}\label{sec:ham}

The (unproximitized) 2DEG is modeled by a single electron band with effective
mass $m^{\ast}$ at electro-chemical potential $\mu$ as described by the
following Bogoliubov-de Gennes Hamiltonian ($e = \hbar = 1$):
\begin{eqnarray}
  H ( x, y) & = & \left( - \frac{\partial_x^2 + \partial_y^2}{2 m^{\ast}} -
  \mu \right) \tau_z \nonumber\\
  &  & - i \alpha ( \sigma_x \partial_y - \sigma_y \partial_x) \tau_z + E_Z
  \sigma_y / 2.  \label{eq:ham}
\end{eqnarray}
In the second line, we added the Rashba spin-orbit coupling (with velocity
$\alpha$) and the Zeeman energy ($E_Z$) due to a magnetic field. The
Hamiltonian acts on the four-component spinor $\tmmathbf{\psi} ( x, y) = [
\psi_{e, \uparrow} ( x, y), \psi_{e, \downarrow} ( x, y), \psi_{h, \downarrow}
( x, y), - \psi_{h, \uparrow} ( x, y)]^T$ containing the electron $( e)$ and
hole $( h)$ components for spin $\sigma = \uparrow, \downarrow$. The Pauli
matrices $\tau_i$ and $\sigma_i$ ($i = x, y, z$) act on particle-hole and spin
space, respectively.

The proximity effect of the superconducting top layer can by included by
integrating out the superconductor in the wide-band limit
{\cite{SauProximityEffect,StanescuProximityEffect,Danon15}}. The resulting
retarded Green's function of the 2DEG is given by
\begin{eqnarray}
  G^R ( x, y, E) & = & \frac{1}{E - H ( x, y) - \Sigma^R_s ( x, y, E) + i
  0_+},  \label{eq:gr}
\end{eqnarray}
with the retarded superconductor self energy
\begin{eqnarray}
  \Sigma_s^R ( x, y, E) & = & p ( E) \Gamma ( x, y) \frac{\Delta \tau_x - E
  \tau_0}{\sqrt{| \Delta^2 - E^2 |}},  \label{eq:sigma}
\end{eqnarray}
and prefactor
\begin{eqnarray}
  p ( E) & = & \left\{ \begin{array}{ll}
    1 & | E | < \Delta,\\
    i \tmop{sgn} ( E), & | E | > \Delta .
  \end{array}  \right.
\end{eqnarray}
In our numerical calculations, we replace $0_+ \rightarrow \eta$ with a
positive $\eta$ smaller than all other energy scales (but we keep the notation
$0_+$ in all following expressions).

The self energy is nonzero only where the device is covered by a
superconducting layer [blue in Figs.~\ref{fig:devices-simple} and
\ref{fig:devices-simulated}] and for simplicity we assume the tunnel coupling
to be uniformly given by a constant $\Gamma$ in these regions
[\Fig{fig:devices-simple}(e) and (f)]. Note that the Zeeman splitting of the
states in the superconducting top layer is neglected because the g factor in
the superconductor ($\approx$2) is smaller than that in the 2DEG (probably
$\approx$10 {\cite{Shabani16}})

Our self energy does not include a proximity-induced shift of the chemical
potential under the superconductor. This approximation is motivated by recent
experiments {\cite{Kjaergaard16}} revealing that proximity-induced
superconductor-normal junctions achieve a high transparency. This indicates
that the mismatch in the chemical potential must be rather small.

Unless stated otherwise, we keep the frequency ($E$) dependence of the self
energy, which leads to a downward renormalization of all non-zero energies \cite{Deng16,vanHeck16}. This affects our estimates of coupling energies and energy
gaps and translates directly into the time scales needed for adiabatic
manipulation. Moreover, the self energy becomes imaginary for energies $\omega
> \Delta$, i.~e., when the states in the 2DEG can also leak into the
superconducting top layer.

\subsection{Tight-binding model and Green's function
formalism}\label{sec:gf}

We simulate the devices numerically by introducing a 2D lattice and applying a
recursive Green's function formalism {\cite{Ferry97Book}} to compute the
density of states in the regions of interest. Since our approach follows that
of {\Cite{Wimmerphd}}, we just sketch its central idea here and refer the
reader to {\Cite{Wimmerphd}} for further details.

First, the discretized versions of the Hamiltonian $H$ and the superconductor
self energy $\Sigma^R_s$ take the form: 
\begin{eqnarray}
  H & = & H_c + H_{c o} + H_o,  \label{eq:h}\\
  \Sigma_s^R & = & \Sigma^R_{s, c} + \Sigma^R_{s, o} .  \label{eq:sr}
\end{eqnarray}
Here, the subscript $c$ denotes the sites in the central region of interest
[surrounded by dashed lines in Figs.~\ref{fig:devices-simple} and
\ref{fig:devices-simulated}], which has $N_x$ ($N_y$) sites in the $x$- ($y$-)
direction. The subscript $o$ denotes all other sites outside. We give concrete
expressions for the discretized versions of $H$ and $\Sigma^R_s$ in App.~\ref{app:tb}. Notably, the self energy is local in space and thus does not
contribute to the coupling of the central and outer region.

To extract MBS coupling energies and topological energy gaps, we compute
the local density of states in the central region as a function of energy $E$.
The local density of states is obtained from the retarded Green's function:
\begin{eqnarray}
  \rho ( n_x, n_y, E) & = & - \tfrac{1}{4 \pi} \tmop{Im} \tmop{Tr} [ G^R_c (
  n_x, n_y, E)] .  \label{eq:locdos}
\end{eqnarray}
Here, the trace runs over particle-hole and spin indices and $n_x$ and $n_y$
refer to the lattice point. The normalization constant has been chosen such
that $\int d E \rho ( n_x, n_y, E) = 1$. We note that the local density of
states below the superconducting gap can be probed directly by tunneling spectroscopy in the limit that the tunnel coupling between probe and 2DEG is much smaller than temperature. All peaks are then broadened by temperature.
To obtain a measure for the excitation spectrum of the central region, we also investigate the total density of states given by
\begin{eqnarray}
  \rho_{\tmop{tot}} ( E) & = & \frac{1}{N_x N_y} \sum_{n_x n_y} \rho ( n_x,
  n_y, E),  \label{eq:totdos}
\end{eqnarray}
where due to the normalization $\int \tmop{dE} \rho_{\tmop{tot}} ( E) = 1$.

The Green's function $G^R_c$ can be found from a Dyson equation derived with
$H_{c o}$ as a perturbation {\cite{Wimmerphd}}:
\begin{eqnarray}
  G^R_c ( E) & = & \frac{1}{E - H_c - \Sigma^R_{s, c} ( \omega) - \Sigma^R_o (
  E) + i 0_+} .  \label{eq:grc}
\end{eqnarray}
Here, $\Sigma^R_o ( E)$ is the self energy describing the effect of the outer
region on the central region. It is nonzero only on the boundary lattice sites
of the central region. Based on {\Cite{Wimmerphd}}, we explain in App.
\ref{app:surfacegreen} how $\Sigma_o^R$ can be obtained from solving a rather
simple eigenvalue problem. Once $\Sigma_{o}^R$ has been determined, one can
in principle compute $G^R_c$ through the inverse in Eq.~(\ref{eq:grc}) to find
the local density of states $\rho$ in the central region. However, since one
actually needs only the the lattice-diagonal entries of $G^R_c$ to compute
$\rho$, one can apply a more efficient recursive technique to determine these
entries. This is further discussed in \App{app:recursive}.

\section{Single Majorana stripe}\label{sec:stripe-single}

To start our analysis of the Majorana devices, we first briefly compare the
topological phase transition and energy spectrum in semi-infinite S- and
N-stripes. We show that a MBS located at the end of a
semi-infinite stripe appears when increasing the magnetic field along the
stripe. We further show that the phase-transition point can be tuned with the
confinement potential, which can be used to couple the MBSs in N- or S-stripes
electrically.

\subsection{S-Majorana stripes}\label{sec:s-stripe}

We first consider an S-type Majorana stripe [\Fig{fig:s-stripe}(a)]. We
assume a top gate that covers the stripe and the region in their vicinity. The
top gate creates a potential well [\Fig{fig:s-stripe}(b)], which is
incorporated in our numerical calculations by lowering the chemical potential
$\mu \rightarrow \mu - \phi_D$ in the unproximitized region [brown in
  \Fig{fig:s-stripe}(b)], while the chemical potential under the superconductor
[blue in \Fig{fig:s-stripe}(b)] is assumed to be unaffected. Such a steep
potential drop is motivated by the good screening effects of the
superconductor. In \App{app:potential}, we verify that a realistically
smoothened gate-induced potential profile has only a minor influence on the
resulting energy spectrum. The potential we use there is a solution of
Poisson's equation assuming zero charge density in the 2DEG. This
approximation is valid in the low charge-density regime considered in
this paper.

\Figure{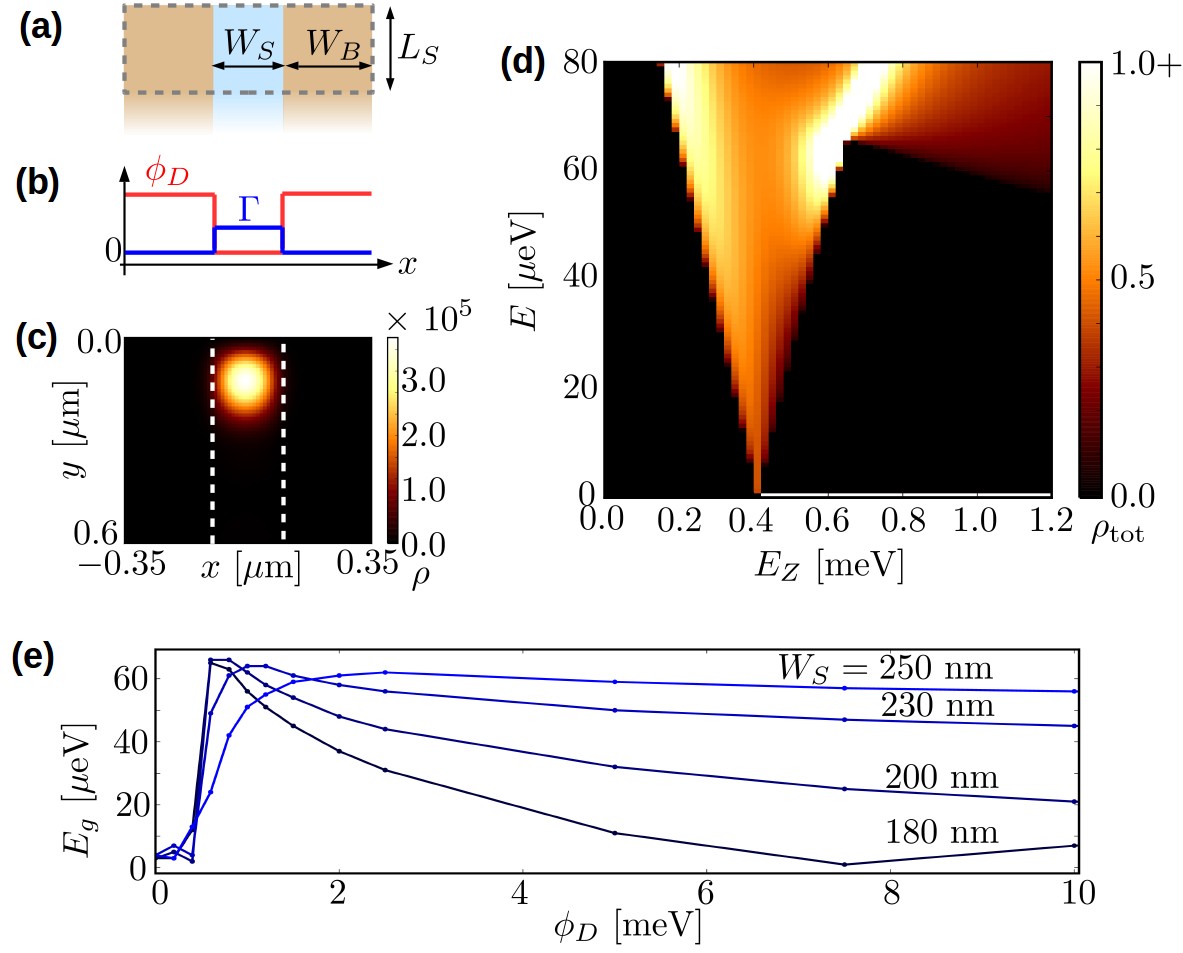}{Majorana bound states and topological phase transition in a semi-infinite S-stripe.
  The device is sketched in (a), and the potential profile (red) is sketched in (b) alongside
  the tunnel coupling to the superconductor (blue). The local density of
  states at energy $E = 0$ is shown in (c) for $E_Z = 0.6~\text{meV}$ and $L_S
  = 600~$nm. The total density of states is shown in (d) for $L_S = 300~\text{nm}$. The potential difference is $\phi_D = 1.5~\text{meV}$ for both
  (c) and (d). The dependence of the lowest excited state energy $E_g$ on the potential barrier height $\phi_D$ is depicted in (e). In all plots,
  we use $W_S = 200~\text{nm}$, $W_B = 500~\text{nm}$, where we show only a
  restricted part in the $x$ direction in (c). The material parameters are $\Gamma
  = 180~\mu \text{eV}$, $\Delta = 235~\mu \text{eV}$, $m^{\ast} = 0.023 m_e$,
  $E_{\tmop{SO}} = m^{\ast} \alpha^2 / 2 = 118.5~\mu \text{eV}$. We further
  chose $\mu = 0$, a broadening $\eta = 10^{- 3}~\mu \tmop{eV}$, and a lattice
  constant of $d = 10~\text{nm}$.\label{fig:s-stripe}}

We next investigate the topological phase transition in an S-stripe by
inspecting the total density of states as a function of energy and Zeeman
energy [\Fig{fig:s-stripe}(d)]. The gap to the energy continuum of states
closes and re-opens at a critical Zeeman energy of $E_Z^{\ast} \sim 420~\mu
\text{eV}$. This value is confirmed by an analytic estimate derived in \App{app:phasetransition}:
\begin{eqnarray}
  E_Z^{\ast} & \approx & 2 \sqrt{\Gamma^2 + \left( \frac{\pi^2}{2 m^{\ast} 
  \tilde{W}^2_S} - \mu - \frac{m^{\ast} \alpha^2}{2} \right)^2} . 
  \label{eq:ezstarmain}
\end{eqnarray}
Here, $\tilde{W}_S$ characterizes the transverse extent of the wave function.
This extent can be estimated by $\tilde{W}_S = W_S + 2 \lambda$, where
$\lambda$ is the decay length into the depleted region. Using the parameters
of \Fig{fig:s-stripe}, and estimating the decay length by $\lambda
\approx 1 / \sqrt{2 m^{\ast} \phi_D}$ (i.e., neglecting magnetic field and
spin-orbit coupling in the estimate), we obtain $E_Z^{\ast} = 423~\mu
\text{eV}$ in good agreement with the numerical result. Note that Eq.
(\ref{eq:ezstarmain}) actually tends to overestimate $E_Z^*$ because it assumes
that the proximity-induced superconductivity with strength $\Gamma$ acts on
the entire extent of the wave function.

\Figure{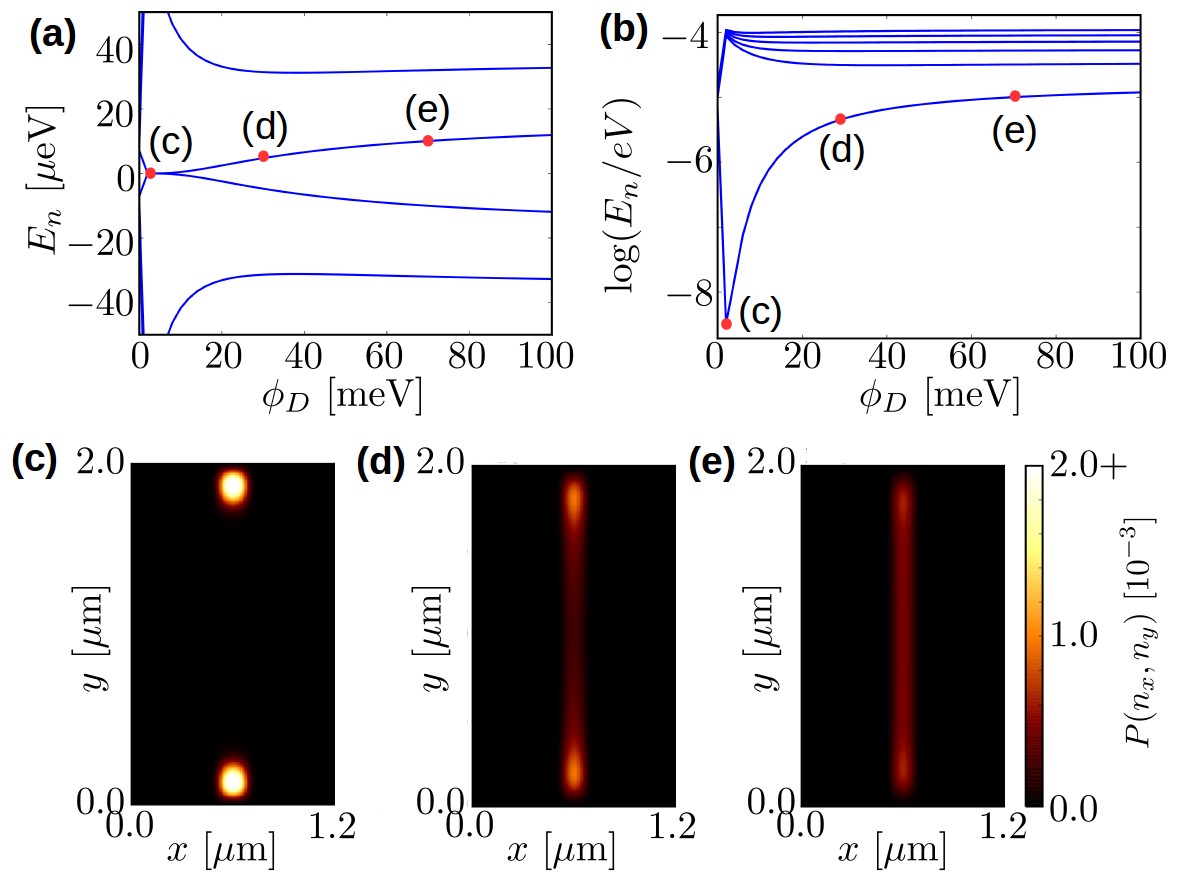}{Gate-tunability
  of \ the MBS wave function overlap in finite S-stripes. Panels (a) and (b)
  show the energy spectrum as a function of the barrier potential
  $\phi_D$ adjacent to the S-stripe [\Fig{fig:s-stripe}(a)] on a
  linear and a logarithmic scale, respectively. In (c)--(d), we show the
  probability density $P ( n_x, n_y)$ for the energy eigenstate closest to
  zero energy for different barrier heights $\phi_D$ as indicated in (a) and
  (b). The stripe length is $L_S = 2~\mu \text{m}$ and all other parameters
   as in \Fig{fig:s-stripe}.\label{fig:s-stripe-control}}

When the Zeeman energies exceeds the critical value, $E_Z > E_Z^{\ast}$, the
density of states exhibits a peak at zero energy [\Fig{fig:s-stripe}(d)].
The corresponding local density of states $\rho ( n_x, n_y)$ at zero energy is
localized at the end of the stripe [\Fig{fig:s-stripe}(c)]. This peak is
due to a MBS that appears at the end of the semi-infinite stripe {\footnote{Our
findings do not strictly prove a topological phase transition at $E_Z =
E_Z^{\ast}$ but this interpretation is in line with other works modeling
nanowires (see, for example, {\cite{Lutchyn11Multiband}}). }}. We can further
see that the topological energy gap -- the energy $E_g$ of the lowest excited
state -- remains stable for $E_Z \gtrsim 600~\mu \text{eV}$ and reaches a
value of about $60~\mu \text{eV}$. This illustrates that a sizable
fraction of the tunnel coupling and the superconducting gap of the parent
superconductor can be reached in an S-type geometry.

Interestingly, the topological energy gap $E_g$ reaches a maximum as a
function of the potential barrier $\phi_D$ [\Fig{fig:s-stripe}(e)]. This feature is related to the tunability of the phase transition point as explained below in this paragraph.
Depending on the stripe width, the topological energy gap may be suppressed or
it may remain nearly constant for increasing potential barrier heights
$\phi_D$. The reason for this is that $\phi_D$ changes the decay length
$\lambda$ of the wave function in the barrier region and thus its transverse
extent $\tilde{W}_S$. From Eq.~(\ref{eq:ezstarmain}), it is clear that this
can tune the system through a phase transition for fixed magnetic field $E_Z$.
Solving Eq.~(\ref{eq:ezstarmain}) for $\tilde{W}_S$ using the same parameters
as before, we find a critical extent of $\tilde{W}^{\ast}_S \approx 260~\text{
nm}$. To estimate the gate-tunability of $\tilde{W}_S$, we first note that
$\tilde{W}_S \rightarrow W_S$ when raising the barriers to large values
$\phi_D \rightarrow \infty$ and $\tilde{W}_S$ is increased by
lowering the barriers. Using a value of $\phi_D = 0.5~\tmop{meV}$, which
is roughly what is needed to form bound states in the stripe
[\Fig{fig:s-stripe}(d)], we obtain $\lambda \approx 60~\text{nm}$. This means
that $\tilde{W}_S$ can be roughly tuned in a range of 120 nm using the
depletion gates.

Our estimates are confirmed by further numerical calculations
[\Fig{fig:s-stripe-control}]. For this purpose, we numerically diagonalized the
Hamiltonian (\ref{eq:ham}) in a finite region with the added zero-frequency
self energy $\Sigma_s^R ( x, y, \omega = i 0) = \Gamma \tau_x$. For simplicity, we neglect finite-frequency corrections here.
For a stripe width $W_S = 200~\text{nm}$, the MBSs split in energy when
increasing the barrier to about 10 meV [\Fig{fig:s-stripe-control} (a)].
We see that the related probability density can be tuned from localized MBSs
[\Fig{fig:s-stripe-control}(c)] over overlapping MBSs wave functions
[\Fig{fig:s-stripe-control}(d)] to an Andreev bound state that is extended along the S-stripe [\Fig{fig:s-stripe-control}(e)].
This means that the MBSs can be controlled electrically: The corresponding
energy of the state can be tuned in a large range between about 1 neV to a
more than 10 $\mu$eV for large barriers [\Fig{fig:s-stripe-control}(b)].
Using a smaller stripe width, one can couple the MBSs for smaller values of
$\phi_D$ [see \App{app:phasetransition}]. However, for separating
adjacent S-stripes from each other, it may be advantageous if the MBS coupling
is suppressed over a larger range of $\phi_D$. We therefore use in the following
calculations $E_Z = 600~\mu \text{eV}$ and $W_S = 200~\text{nm}$ for S-stripe
devices.

\subsection{N-Majorana stripes}\label{sec:n-stripe}

\FigureBottom{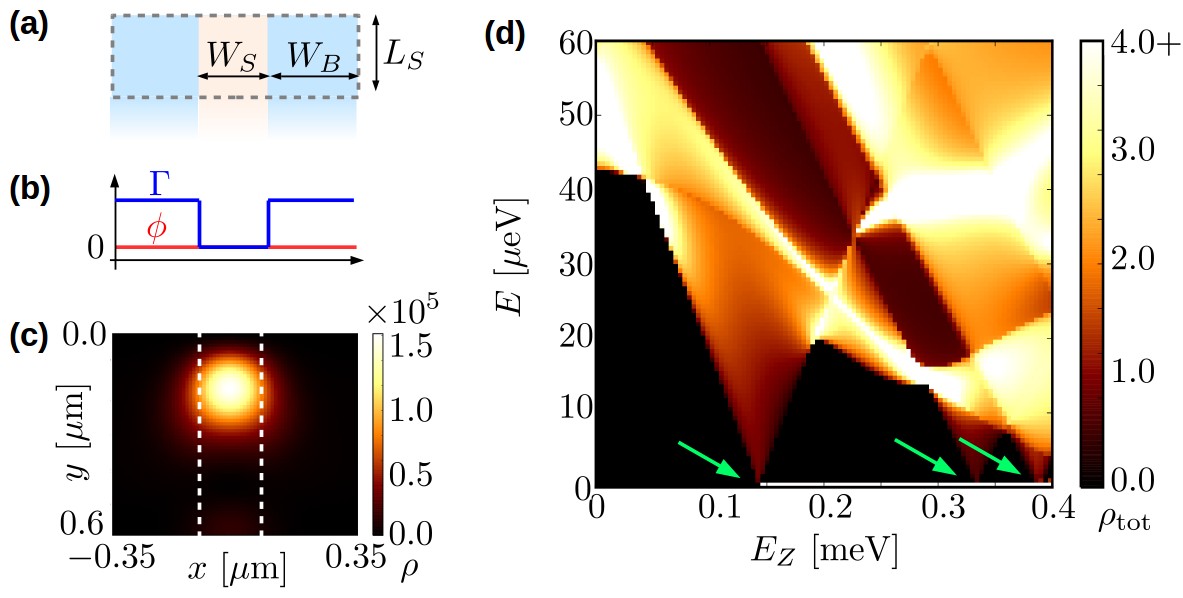}{Majorana
  bound states and topological phase transition in a semi-infinite N-stripe.
  The device is sketched in (a), and the potential profile (red) is sketched in (b) alongside
  the tunnel coupling to the superconductor (blue). The local density of
  states at energy $E = 0$ is shown in (c) for $E_Z = 0.2~\tmop{meV}$ and $L_S
  = 600~$nm. The total density of states is shown in (d) for $L_S = 300~\text{nm}$.
  The green arrows indicate the positions of the gap closings (see text).
  All parameters are as in \Fig{fig:s-stripe}.\label{fig:n-stripe}}

 \Bigfigure{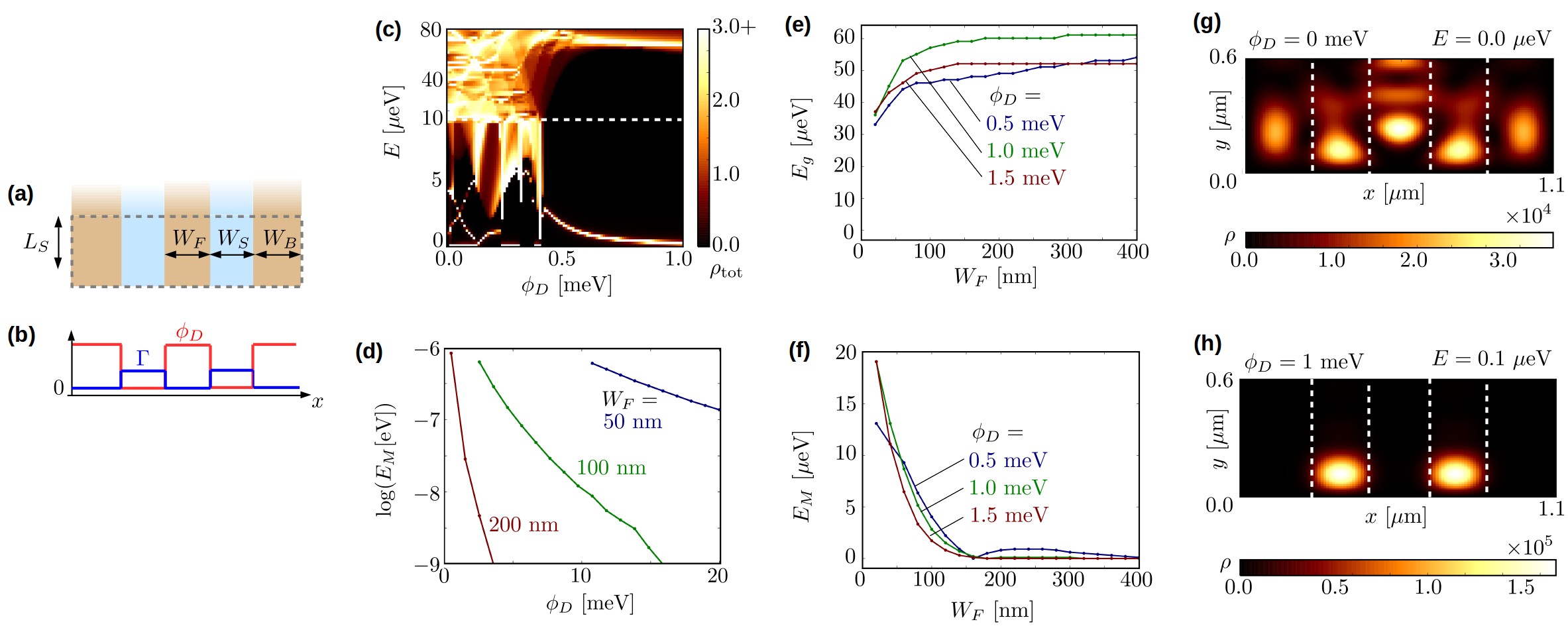}
                    {Suppression
  of MBS coupling in parallel S-stripes with potential-barrier height
  $\phi_D$ in between and stripe distance $W_F$. The device is sketched in (a) and the electrostatic confinement
  potential $\phi ( x)$ (red) is sketched alongside the tunnel coupling $\Gamma (x)$ to
  the superconductor (blue) in (b). The total density of states $\rho_{\text{tot}}$ is mapped out in (c) and the energy $E_M$ of the first
  peak of $\rho_{\text{tot}}$ is depicted in (d) and (f).
  We show the energy $E_g$ of the onset of the energy continuum in (e) and
  the local density of
  states $\rho$ in (g) and (h) for indicated energies $E$ and barrier heights
  $\phi_D$. 
  The Zeeman energy is $E_Z = 0.6~\text{meV}$, and the geometrical
  dimensions are, unless stated otherwise, $L_S = 200~\text{nm}$ [(c)--(f)] / $600~\text{nm}$ [(g) and (h)], $W_S = 200~\text{nm}$, $W_B = 500~\text{nm}$,
  and $W_F = 200~\text{nm}$. We use $\eta = 1~\text{neV}$ except for (f), where $\eta = 0.1~\text{neV}$.
  All other parameters are as in \Fig{fig:s-stripe}.\label{fig:s-stripes-parallel}}

We next compare the spectral properties of S-stripes with those of N-stripes
[\Fig{fig:n-stripe}(a)]. Since single N-stripes have been thoroughly
discussed before {\cite{Hell17a,Pientka16}}, we keep our discussion here
brief. For a discussion of the gate control of the MBS coupling in N-stripes
see {\Cite{Hell17a}}.

Similar to S-stripes, we also find a closing and a re-opening of the energy
gap at about $E_Z^{\ast} \sim 140~\mu$eV [\Fig{fig:n-stripe}(d), left green arrow]. Again,
a zero-energy peak appears in the total density of states with a local density
[\Fig{fig:n-stripe}(c)] similar to that for S stripes. This MBS
peak even persists when the gap closes and re-opens a second and a third time at
$E_Z^{\ast} \sim 320~\mu$eV and $E_Z^{\ast} \sim 390~\mu$eV, respectively [\Fig{fig:n-stripe}(d), middle and right green arrow]. This is probably due to to co-existence of two or three
MBSs per end, which is a consequence of the additional spatial mirror symmetry
of the device, which is therefore in symmetry class BDI
{\cite{Hell17a,Pientka16}}.

However, there are also differences in the characteristics of S- and
N-stripes. First, the critical Zeeman field in N-stripes, $E_Z^{\ast} \sim 140~\mu$eV,
is {\tmem{below}} 2$\Gamma$ and thus smaller than that for S-stripes,
which is expected to be larger than 2$\Gamma$. As explained in Ref.
{\cite{Hell17a}}, the lower value of $E_Z^{\ast}$ is due to the weakened
superconducting proximity effect in the normal region. This is an advantage
for reaching the topological regime as one strives to minimize the required
magnetic fields. However, the magnetic field range that results in a sizable
energy gap to excited states is smaller than that in S-stripes. Furthermore,
the energy gap to the excited states is smaller (about 20 $\mu$eV at $E_Z =
200~\mu$eV) than that for S-stripes. However, it can be increased by asymmetric
N-stripe designs as shown numerically {\cite{Hell17a}} (see also \Fig{fig:n-stripes-parallel}). This shows that engineering MBSs in N-stripes
should also be possible but requires more optimization than in S-stripes.

\section{Parallel Majorana stripes}\label{sec:stripe-parallel}

To build more complex Majorana devices or even Majorana networks, it is
important to specify how close two Majorana stripes can be placed next to each
other. The coupling energy of the MBSs from the two stripes should be
suppressed as much as possible, while the energy gap to excited states should
remain as large as possible. Considering next two Majorana stripes in parallel
[Figs.~\ref{fig:devices-simulated}(a) and (b)], we show that the MBSs are well
separated for stripe distances of about 200 nm and moderate potential
differences on the scale of a few meV. This geometrical constraint is also
relevant for the design of a trijunction in a tuning-fork geometry discussed
in \Sec{sec:tuning-fork}.

\subsection{S-Majorana stripes}\label{sec:s-stripes-parallel}

   \Bigfigure{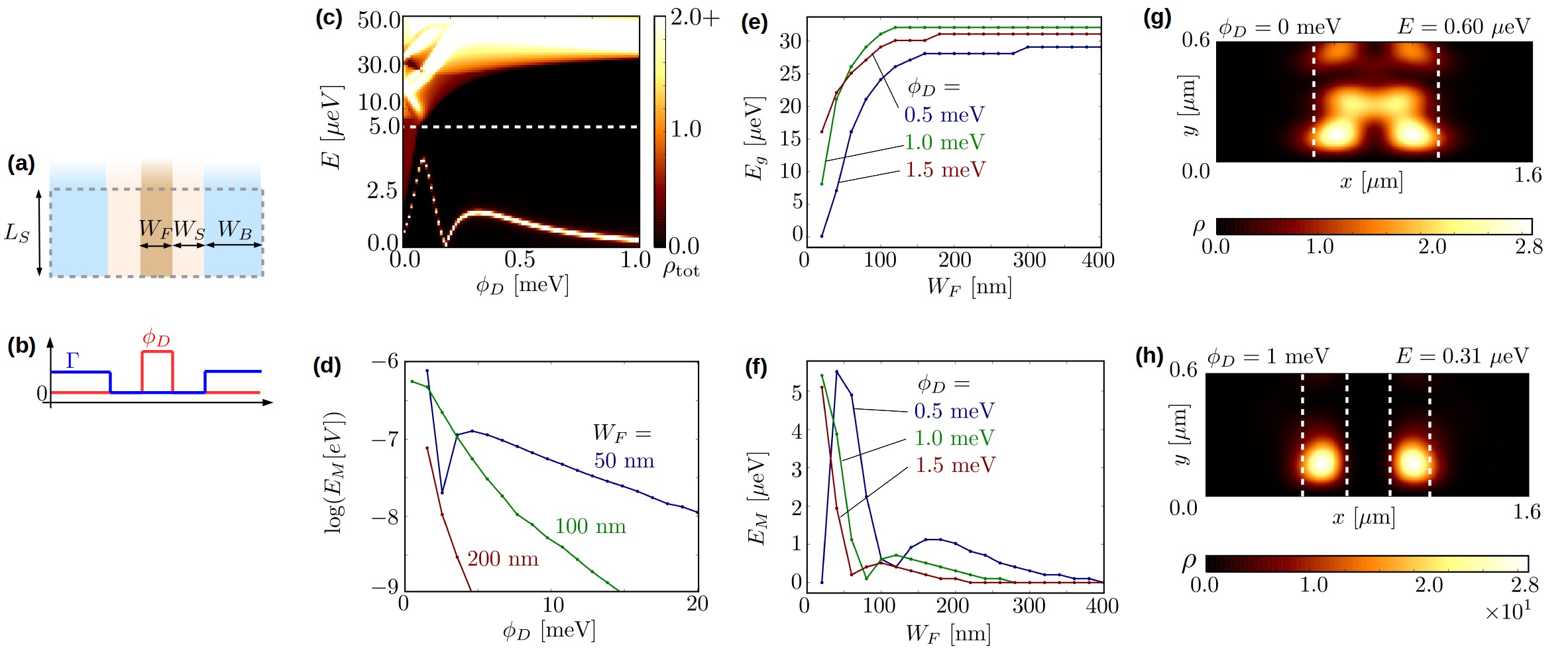}
                                   {Suppression
  of MBS coupling in parallel N-stripes with potential-barrier height
  $\phi_D$ and stripe distance $W_F$. The device is sketched in (a) and the electrostatic confinement
  potential $\phi (x)$ (red) is sketched alongside the tunnel coupling $\Gamma ( x)$ to
  the superconductor (blue) in (b). The total density of states $\rho_{\text{tot}}$ is mapped out as a
  function of both these parameters in (c). The energy $E_M$ of the lowest
  peak of $\rho_{\text{tot}}$ is given in (d) and (f) and we show the energy $E_g$ of the onset of the energy continuum in (e).
  We show the local density of
  states $\rho$ in (g) and (h) for indicated energies $E$ and barrier heights
  $\phi_D$.
  We use the same parameters as in \Fig{fig:s-stripes-parallel} except for a
  different Zeeman energy of $E_Z = 0.2~\text{meV}$.
  \label{fig:n-stripes-parallel}}

We first discuss two parallel S-stripes
[\Fig{fig:s-stripes-parallel}(a)], where a single top gate covers both stripes.
As before, we assume that the gate voltage shifts the chemical potential $\mu
\rightarrow \mu - \phi_D$ only where the 2DEG is not covered with a
superconductor [\Fig{fig:s-stripes-parallel}(b)].

The potential-barrier dependence of the total density of states is depicted in
\Fig{fig:s-stripes-parallel}(c). For the parameters used in this case,
the density of states changes drastically around $\phi^{\ast}_D \approx 0.4
~\text{meV}$.
Below this value, the density of states is nonzero nearly in the
entire energy range and the local density of states is not localized in the
stripe regions [\Fig{fig:s-stripes-parallel}(g)]. The lower edge of the
continuum of states is suppressed to less than $10~\mu$eV and we find a few
discrete states below that. The small remaining energy gap to the continuum is
probably due to the finite width of the device in direction perpendicular to
the stripes. This leads to a small confinement energy of $\pi^2 / [ 2 m^{\ast}
( 2 W_B + 2 W_S + W_F)^2] \approx 8~\mu \text{eV}$.

By contrast, for potential barriers $\phi_D$ above the threshold
$\phi^{\ast}_D$, the regions without top superconductor are depleted. While this threshold certainly depends on the
parameters of the system, it corresponds to an energy comparable to the
induced superconducting gap and the spin-orbit energy and we therefore expect
it to hold more generally. The density of states has a peak approaching zero
energy that persists for all $\phi_D > \phi_D^{\ast}$ in the range shown. This peak can be attributed to two slightly overlapping MBS whose local density of states is confined to the stripe region
[\Fig{fig:s-stripes-parallel}(h)].

The coupling between the MBSs can be efficiently suppressed by increasing the
potential barrier or the separation of the stripes.
We determine the coupling energy $E_M$ of neighboring MBS as the position of the
first peak in the total density of states [extraction procedure explained 
in \Fig{fig:s-stripes-parallel}]. Specifically, we find that for a stripe
separation of $W_F = 200~\text{nm}$, a potential barrier $\phi_D$ of only a
few meV is needed to suppress the Majorana coupling $E_M$ to the neV range
[\Fig{fig:s-stripes-parallel}(d)]. We further find that $E_M$ stays below about 1~$\mu$eV for stripe separations of about
$W_F > 200~\text{nm}$ for potential barriers as low as $0.5~\text{meV}$ [\Fig{fig:s-stripes-parallel}(f)].
However, to reach the neV range for separations much smaller than 100 nm requires much
larger potential barriers [\Fig{fig:s-stripes-parallel}(d)].

We next investigate the behavior of the energy $E_g$ of the lowest excited states. We can see that for potential barriers $\phi_D > \phi_D^{\ast}$, the threshold of the continuum is pushed up to about $E_g \sim 60~\mu$eV [\Fig{fig:s-stripes-parallel}(c)]. Furthermore, $E_g$ increases with the separation $W_F$ and clearly depends
on the height of the potential barrier [\Fig{fig:s-stripes-parallel}(e)], similar to what has been found for the
single S-stripes [\Fig{fig:s-stripe}(e)].  We checked that even for larger values $\phi_D$, as used in
\Fig{fig:s-stripes-parallel}(d), the topological gap remains large (The intra-stripe coupling of the MBS is ignored here, compare \Fig{fig:s-stripe-control}).

\subsection{N-Majorana stripes}\label{sec:n-stripes-parallel}

We next discuss a configuration of two N-stripes in parallel
[\Fig{fig:n-stripes-parallel}(a)]. Here, we consider a gate which depletes a
non-superconducting region between the two stripes
[\Fig{fig:n-stripes-parallel}(b)]. We show next that the N stripes can be
separated with similar parameter choices as for the S stripes.

First, at zero potential difference, $\phi_D = 0$, the 2DEG region without top
superconductor forms one wider stripe and the total density of states is
nearly gapless. However, the density of states is small at low energies [dark
red in the color scale of \Fig{fig:n-stripes-parallel}(c) for small
$\phi_D$]. The corresponding local density of states spreads over the entire
2DEG region without top superconductor [\Fig{fig:n-stripes-parallel}(g)].

For nonzero potential barrier, the lower edge $E_g$ of the continuum of
excited states raises quickly to higher energies. The total density of states
features furthermore a discrete peak at energy $E_M < E_g$, which approaches
zero energy [\Fig{fig:n-stripes-parallel}(c)]. The local density of
states at this peak is clearly localized at the ends of both stripes
[\Fig{fig:n-stripes-parallel}(h)], i.e., the wider stripe is cut into two when
increasing $\phi_D$. We interpret the energy $E_M$ again as the coupling
energy of the MBSs localized at the ends of the two stripes. Similar to
S-stripes, this coupling energy can be strongly suppressed by increasing
$\phi_D$ [\Fig{fig:n-stripes-parallel}(d)]. Again, $E_M$ reaches the
sub-neV range for stripe separations of $W_F=$200 nm and $\phi_D$ of a few
meV. We further find that the Majorana coupling energy $E_M$ is suppressed when
increasing the stripe distance $W_F$
[\Fig{fig:n-stripes-parallel}(f)]. By comparing these results to that
for S-stripes, it seems that similar potential barriers and stripe distances
are needed to ensure a good separation in the case of N-stripes.

We finally investigate the parameter dependence of the lower edge $E_g$ of the
excited-state continuum. The energy $E_g$ reaches a value around 20 $\mu$eV for
potential differences around 1 meV
[\Fig{fig:n-stripes-parallel}(e)], so that the barrier exceeds all
other energy scales. This value is somewhat larger than what we found for for
the single-stripe case [\Fig{fig:n-stripe}]. This is consistent with Ref.
{\cite{Hell17a}}, which showed that an asymmetric design of a single N stripe
with a proximity-induced superconductor on one side and a gate on the other
leads to a larger topological gap than a device with superconducting regions
on both sides. Yet, the energy gap $E_g$ for N-stripes is smaller than that
for S-stripes, which is expected due to the weaker superconducting proximity
effect for N stripes (see \Sec{sec:n-stripe}).

\section{Trijunctions: Tuning-fork design}\label{sec:tuning-fork}

\Bigfigure{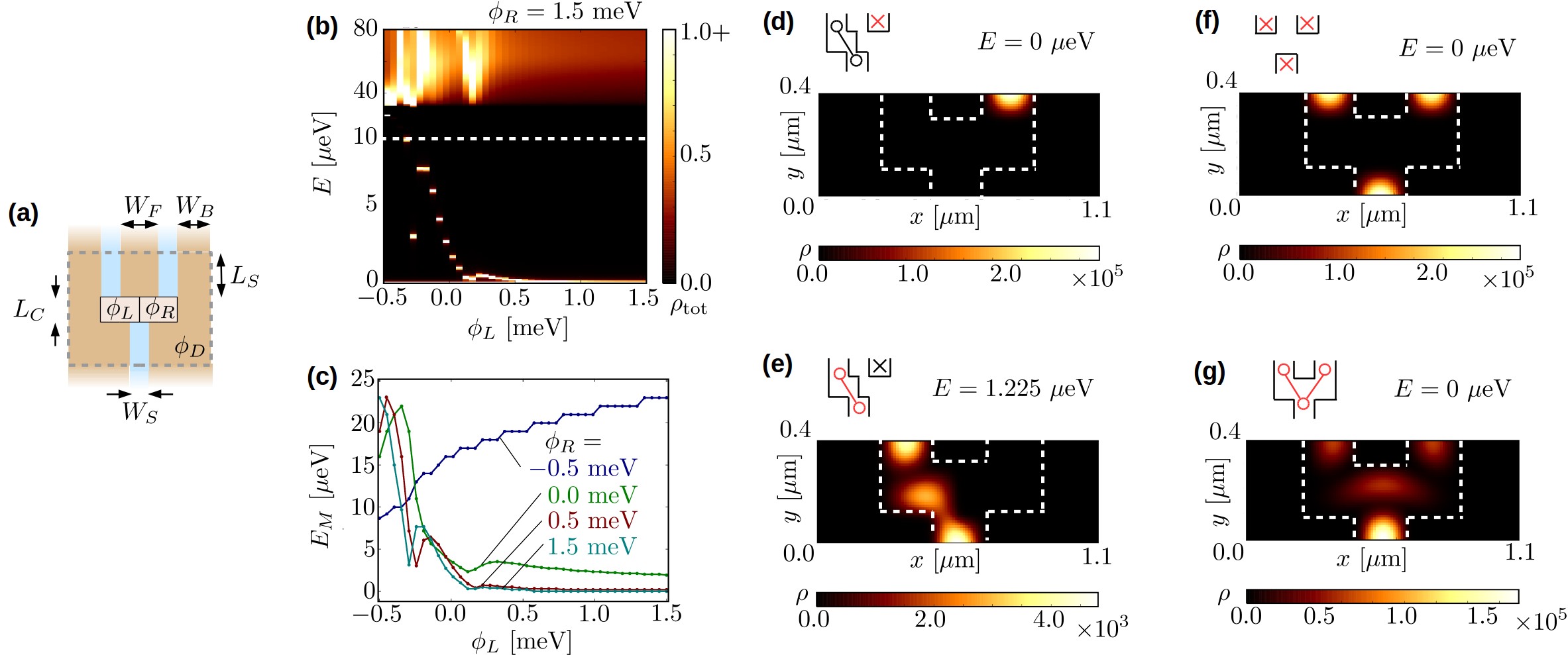}{Controlling
  the coupling of three MBSs in a tuning-fork trijunction formed from
  S-stripes. The device is sketched in (a). The total density of states
  $\rho_{\tmop{tot}} ( E)$ is shown in (b) for fixed
  depletion potential $\phi_R = 1.5~\text{meV}$ at the right junction as a
  function of the potential difference $\phi_L$ at the left junction.
  We compute $\rho_{\text{tot}}$ for energies in steps of $0.1~\mu\text{eV}$ ($1~\mu\text{eV}$) below (above) $10~\mu\text{eV}$ and for potential values $\phi_L$ in steps of $0.05~\text{meV}$. This finite resolution leads to the step-like changes of the subgap peak position.
  In (c),
  we show the energy $E_M$ of the first peak of $\rho_{\tmop{tot}} ( E)$ at
  nonzero energy for different values of $\phi_R$ as indicated.
  Panels (d)--(g) show for different coupling configurations (sketched in top part) the local density of states $\rho ( E, n_x, n_y)$ in the central region for
  different potential barrier heights (bottom part): (d) and (e) $\phi_L = 0~\tmop{meV}$,
  $\phi_R = 5~\text{meV}$, (f) $\phi_L = \phi_R = 5~\text{meV}$, and (g)
  $\phi_L = \phi_R = 0~\tmop{meV}$. The energies $E$ are indicated in the
  figure. For illustrational purposes, we chose the maximal values of $\phi_L$
  are $\phi_R$ to be larger than in (b) and (c). The finite-energy
  peak in (e) is thus shifted to a smaller value (which could be increased by
  lowering $\phi_R$).
  In all plots, the depleted
  areas are at a potential $\phi_D = 5~\text{meV}$, the Zeeman energy is $E_Z
  = 0.6~\text{meV}$, and the dimensions of the device are given by $L_S = 100~\text{nm}$,
  $L_C = 200~\text{nm}$, $W_F = 200~\text{nm}$, $W_S = 200~\text{nm}$,
  $W_B = 500~\text{nm}$. We use $\eta = 1~\text{neV}$ and all
  other parameters are as in \Fig{fig:s-stripe}.\label{fig:s-fork}}

We next discuss the design of a trijunction using Majorana stripes [Figs.~\ref{fig:devices-simulated}(c) and (d)]. An important design constraint
discussed in {\Cite{Hell17a}} is that all Majorana stripes have to placed
in parallel {\footnote{This has been discussed for N-stripes in the
    supplemental material of {\Cite{Hell17a}}. 
}}. The reason is that the topological
energy gap closes when the magnetic field direction deviates from the long
stripe direction by more than about 10 degrees. This is why we investigate a
tuning-fork shaped structure as sketched in Figs.~\ref{fig:devices-simulated}(c) and (d).
This raises the crucial question whether topologically trivial low-lying
excited states occur in the central coupling region, which would be
detrimental to the operation of this Majorana device. We show here for
specific examples for both S- and N-type designs that this can be avoided while
controlling the MBS coupling energies between $\sim 1$ neV and $\sim 10~\mu$eV using electric gates. The energy gap to the excited states remains
throughout larger than 10 $\mu$eV.

\subsection{S-Majorana stripes}\label{sec:s-fork}

We first investigate an S-stripe trijunction
[\Fig{fig:s-fork}(a)]: The 2DEG regions without a superconducting top
layer are depleted by a top gate except for a central region connecting the
three stripes. In this region, two additional gate 'fingers' with different
voltages are integrated, which are isolated from each other. They affect only
the central part by coupling the left and right upper stripe to the lower
stripe. Such gates could be fabricated, for example, by adding a second
metallization layer on top the superconducting film and the depletion gate.
For the concrete devices studied in this section, these two gate 'fingers'
would have a width of 200 nm, which should be achievable with present-day
fabrication technologies.

To discuss the voltage control of the MBS coupling, we start from the
situation when the channel connecting the right and lower stripe is depleted
($\phi_R = 1.5~\text{meV}$), while the gate controlling the channel connecting
the left and the lower stripe is changed. When $\phi_L$ is around zero or below, we
can discern two discrete peaks below the threshold to the continuum of states
[\Fig{fig:s-fork}(b)]. One of the peaks is at zero energy [hardly visible in \Fig{fig:s-fork}(b)] and
the corresponding local density of states is localized at the end of the right
stripe [\Fig{fig:s-fork}(d)]. We interpret this feature as a MBS
at the end of the right stripe, which cannot couple to the other MBS in this
gating configuration. The second peak is at a few $\mu$eV and the
corresponding local density of states is delocalized between the lower and the
left stripe [\Fig{fig:s-fork}(e)]. We attribute this peak to a
coupling of the two MBSs at the end of the lower and left stripe. They form a
fermionic mode at energy $E_M$. When lowering $\phi_L$ to $-0.5~$meV, the
coupling energy increases up to about 20~$\mu\text{eV}$ [\Fig{fig:s-fork}(c)].

The energy of the mode formed by the two coupled MBSs can be controlled by
the left gate [\Fig{fig:s-fork}(b) and (c)]. When increasing the
potential height $\phi_L$, the energy of this mode approaches zero. For
$\phi_L \gtrsim 1~\tmop{meV}$, the two MBSs are separated and the local
density of states at zero energy shows three peaks at each end of the three
stripes [\Fig{fig:s-fork}(f)]. The peak energy reaches a value
below 1 neV at $\phi_L = 5~$meV (not shown).

Moreover, the lower edge of the continuum of excited states stays rather
constant at about $E_g \approx 30~\mu\text{eV}$ when increasing $\phi_L$
[\Fig{fig:s-fork}(b)]. This means that one can can tune the
coupling between the left and lower MBSs without introducing low-lying
excitations and keep the right MBS separated. The value for $E_g$ is somewhat
reduced as compared to Figs.~\ref{fig:s-stripe} and
\ref{fig:s-stripes-parallel} because we use a larger depletion potential
$\phi_D$ here, which moves the system closer to the phase-transition point.
This larger value of $\phi_D$ reduces the MBS coupling between the parallel
Majorana stripes. This is needed to reach coupling energies on the neV range as discussed in \Sec{sec:s-stripes-parallel}.

Finally, by lowering the potential $\phi_R$ on the right part, the right MBS
can couple to the other ones. At $\phi_R = \phi_L = 0$, the three MBSs form
one MBS with a zero-energy peak in the density of states. The corresponding
local density of states is delocalized over all three stripes
[\Fig{fig:s-fork}(g)]. In addition, the MBSs form a fermionic mode
leading to a peak in the density of states at a finite energy $E_M$.
The energy of this mode can be increased above 10 $\mu$eV by lowering $\phi_R$
to negative values [\Fig{fig:s-fork}(c)]. As $\phi_L$ is
increased and the left MBS is decoupled, $E_M$ even increases.
We checked that the lowest value of further excited states stays rather
constant at about 30~$\mu\text{eV}$ also for reduced $\phi_R$, similar to
\Fig{fig:s-fork}(b). This implies that one can also tune between
three and two coupled MBSs without introducing low-lying excited states.

\subsection{N-Majorana stripes}\label{sec:n-fork}

We finally discuss a trijunction formed by three N-stripes
[\Fig{fig:n-fork}(a)]. We show that one can gain voltage control over
the MBSs in an N-stripe trijunction very similar to the S-stripe trijunction
and compare the device operation and coupling energies for both designs.

We consider again a tuning-fork shaped device similar to that for S-stripes
[\Sec{sec:s-fork}]. The geometrical dimensions of both devices considered
differ only in that the separation of the two upper stripes is slightly larger
for the N-stripes. The finger gates again only control the central part.
However, one has to deplete the regions left and right to the central part in
addition to the region between the two upper N-stripes. Alternatively, one
could disconnect the different parts by etching the 2DEG. Separating the three
Majorana stripes in this way is crucial for the proper operation of the
device: Just replacing these parts by a superconducting top layer is not
sufficient as discussed below in this section.

\Bigfigure{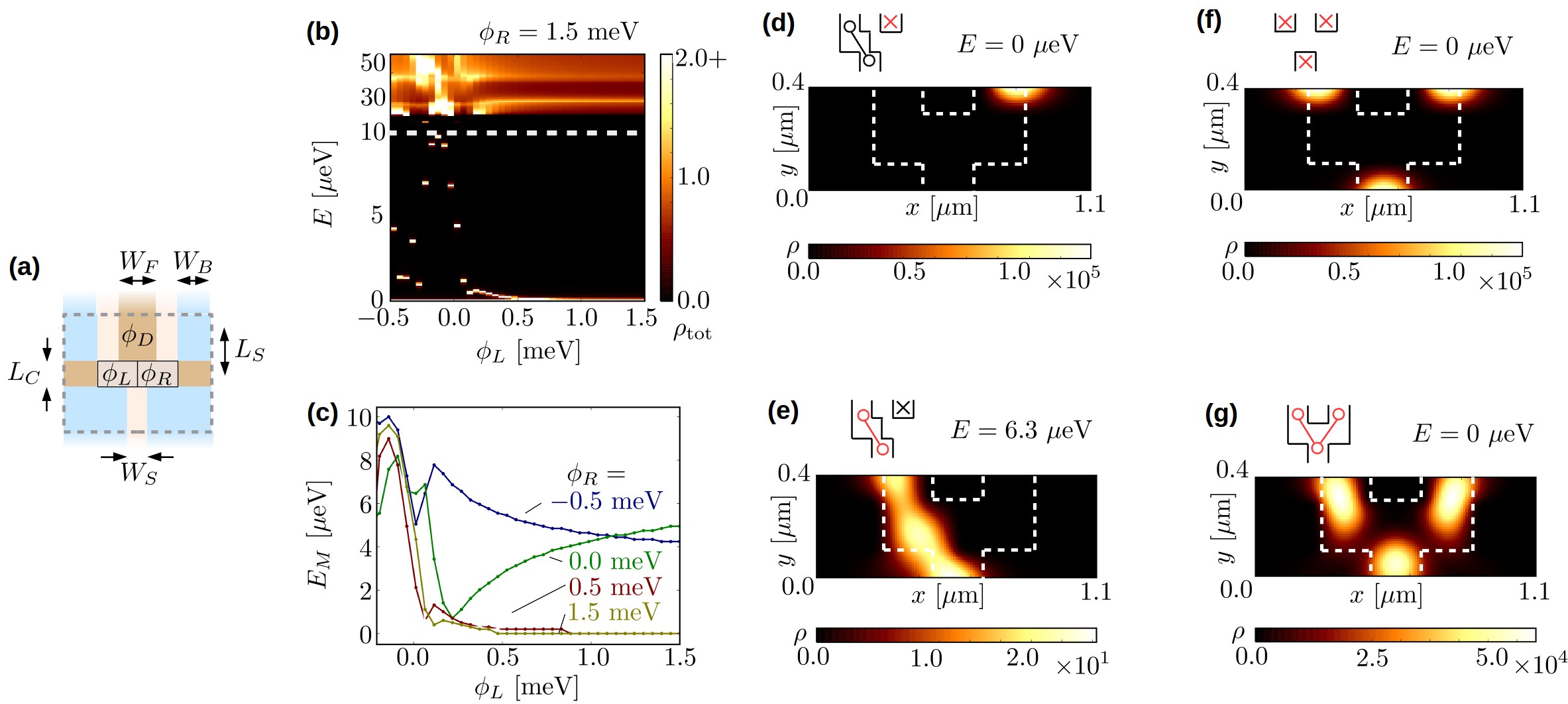}{Controlling
  the coupling of three MBSs in a tuning-fork trijunction formed from
  N-stripes. The device is sketched in (a). The total density of states $\rho_{\tmop{tot}}
  (E)$ is shown in (b) for potential barrier height
  $\phi_R = 1.5~\text{meV}$ at the right junction as a function of the
  potential $\phi_L$ applied to the left junction. In (c), we show the energy
  $E_M$ of the first peak of $\rho_{\tmop{tot}} (E)$ at nonzero energy for
  different values of $\phi_R$ as indicated (extraction procedure discussed in \Fig{fig:s-fork}).
  In the bottom part of the panels (d)--(g), we show the
  local density of states $\rho ( E, n_x, n_y)$ in the central region for
  different potential barrier heights: (d) and (e) $\phi_L = 0~\tmop{meV}$,
  $\phi_R = 5~\text{meV}$, (f) $\phi_L = \phi_R = 5~\text{meV}$, and (g)
  $\phi_L = \phi_R = 0~\tmop{meV}$. The energies $E$ are indicated in the
  figure. We use different maximal values for $\phi_L$ and $\phi_R$ as
  compared to (b) and (c) for illustrational purposes.
  We chose $\phi_D = 5~\text{meV}$, $E_Z = 0.2~\tmop{meV},$ $W_F = 250~$nm, and all other parameters are
  as in \Fig{fig:s-fork}.\label{fig:n-fork}}

We start our discussion of the gate control again for the situation when the
right stripe is decoupled from the lower and the left one by choosing $\phi_R
= 1.5~$meV. The corresponding total density of states has a zero-energy peak
for all values of $\phi_L$ [\Fig{fig:n-fork}(b)]. In analogy to
the S-stripe trijunction, we find that the local density of states is
localized in the right stripe for $\phi_L = 0$
[\Fig{fig:n-fork}(d)], which corresponds again to a decoupled MBS in the
right stripe {\footnote{In contrast to the S-stripes, we can see that the
local density of states is not centered in the middle of the upper N-stripes
The reason is the asymmetric device design in this case: The MBSs can extend
somewhat into the region with proximitized-induced superconductivity, while
they cannot penetrate the depleted region in the middle between the upper
stripes.}}. The two MBSs in the lower and left stripe are coupled and lead to
a peak in the density of states at finite energy $E_M$
[\Fig{fig:n-fork}(b)]. At the peak energy $E_M$, the local density of
states is indeed distributed over both stripes and the central region
[\Fig{fig:n-fork}(e)]. By increasing $\phi_L$, the discrete peak $E_M$
in the total density of states approaches zero energy
[\Fig{fig:n-fork}(b)]. It reaches a value of about 0.5 $\tmop{neV}$ at
$\phi_L = \phi_R = 5~$meV (not shown). The local density of states at zero
energy shows three disconnected peaks in this case, corresponding to three
isolated MBSs [\Fig{fig:n-fork}(f)].

If we next lower $\phi_R$ to $- 0.5~\tmop{meV}$, we can see that the peak
energy $E_M$ does not drop to zero as $\phi_L$ is increased
[\Fig{fig:n-fork}(c)] but stays at about $\sim 5~\mu$eV, somewhat smaller than
for S-stripes. In addition, a zero-energy peak remains in the total
density of states throughout, which is not shown here. This peak corresponds
to a MBS formed from a superposition of the three MBSs in the three stripes as
the local density of states illustrates [\Fig{fig:n-fork}(g)].

We note that the gate control works properly only if the two regions left
and right of the central coupling region are depleted. If the depletion gates
there are replaced by a superconducting top layer (but keeping device the same
otherwise), we find a peak energy of about $E_M \gtrsim 0.3~\mu$eV (not shown
here). This peak energy cannot be suppressed further by increasing the
potential barriers $\phi_L$ or $\phi_R$. This is consequence of an undesired
coupling of the MBSs in the upper and lower stripe through the region with
proximity-induced superconductivity. The barrier induced by the superconductor
(here about 0.2 meV) is not enough to separate the stripes sufficiently.

One could also imagine that the potential in the depleted regions left and
right of the center region was also controlled with the gates. This would have
the advantage that only a single gate was needed on each side of the trijunction.
However, one loses good control over the MBS coupling energies in this case.
We checked from numerical calculations (not shown) that in this case the
energy gap $E_g$ to the excited states is suppressed for low $\phi_L$ and
$\phi_R$ when the MBS coupling becomes sizable. One may explain this feature
by considering that the confinement energy in the direction perpendicular to the
wires is decreased in this case. One then essentially creates a channel for
electrons perpendicular to the magnetic field. From our results in Ref.
{\cite{Hell17a}}, we know that the topological energy gap closes when the
magnetic field is perpendicular to the stripe.

Finally, our numerical calculation shows that there remains a gap to the
continuum of excited states [\Fig{fig:n-fork}(b)], also when $\phi_R$ is
reduced (not shown). However, we can see that this energy gap is smaller than
that for the S-stripe design: For all potentials $\phi_L$, the continuum of
excited states appears at $E_g \sim 20~\mu$eV, a value that is comparable to
the value in \Fig{fig:n-stripe} and most probably limited by the
symmetric design of the lower Majorana stripe. We expect that $E_g$ could be
increased somewhat by an asymmetric device geometry for the lower stripe (see
\Sec{sec:n-stripes-parallel}).

\section{Time scales for braiding}\label{sec:braiding}

Based on our numerical results from above, we finally discuss the time scales
for braiding MBSs in the two trijunction designs. The protocol we consider is
sketched in \Fig{fig:braiding} and relies on adiabatic manipulation of
the coupling energies of the MBSs {\cite{Aasen16,SauBraiding}}. In addition to
the control of the MBS coupling in the central region of the trijunction,
braiding also requires control over the MBS coupling within a Majorana stripe,
both for initialization and readout as well as the braiding steps themselves.
Before turning to the time scales in \Sec{sec:timescales}, we first
discuss the intra-stripe coupling strategies in \Sec{sec:intrastripe-coupling}.

\Figure{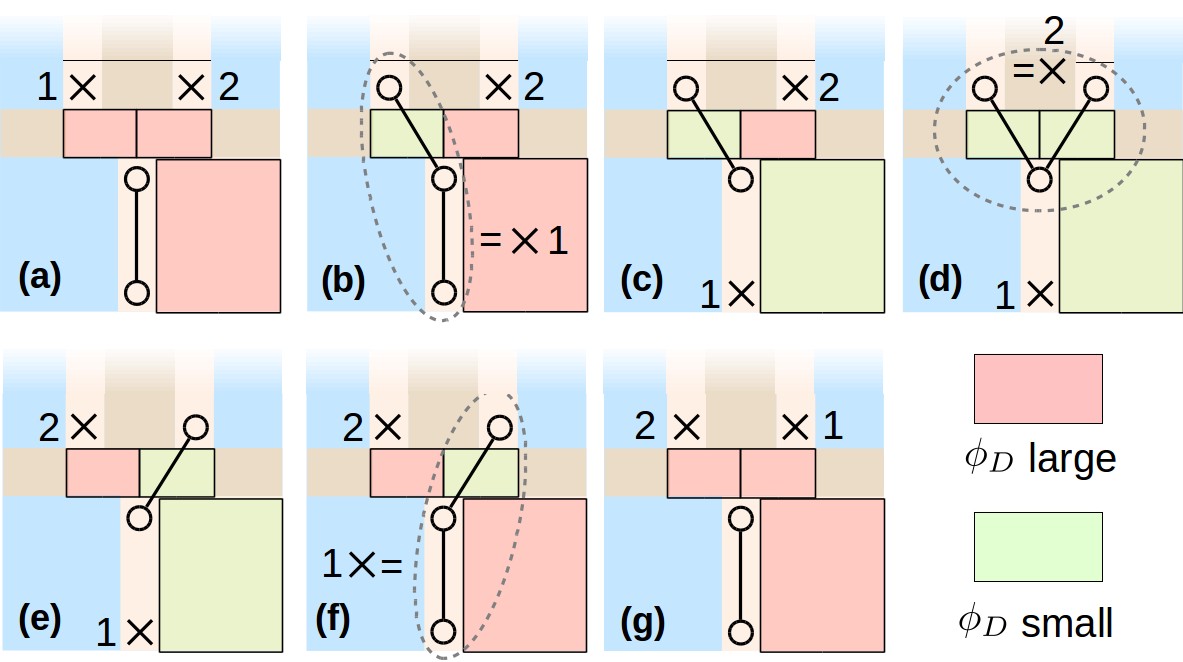}{Gate-controlled
  adiabatic braiding scheme. The steps shown in (a)--(g) exchange the MBSs 1
  and 2 located initially at the lower ends of the two upper stripes. The
  crosses indicate MBSs at zero energy, while the circles and lines indicate
  fermionic modes at finite energy. Note that coupling three MBSs results in
  one fermionic mode at finite energy and one MBS at zero energy, which is
  delocalized between the positions of the three MBSs when uncoupled [as in
  (b), (d) and (f)]. Even though the protocol is here illustrated for
  N-stripes, the protocol can be carried out in the same way also for S-stripe
  trijunctions.\label{fig:braiding}}

\subsection{Gate-voltage control over intra-stripe Majorana bound state
couplings}\label{sec:intrastripe-coupling}

The MBS coupling within the Majorana stripes can be tuned in two ways, namely
by (i) tuning the Majorana wave function overlap, or by (ii) tuning the
Coulomb interaction in the wire.

The first strategy can be achieved by a gate-controlled tuning of the
effective stripe width as discussed for S-stripes in \Sec{sec:s-stripe}
and for N-stripes in {\Cite{Hell17a}}. In both cases, one can tune the
MBS coupling energies between about 1 neV or less to about $E_M = 10~\mu$eV or
larger, while maintaining a gap to the excited states larger than $10~\mu$eV.
This is comparable to the results for the trijunction region.

The second strategy to gain control over the MBS coupling would be to take
advantage of Coulomb blockade physics, similar to proposals based on nanowires
{\cite{Aasen16,Hell16}}. The idea is to electrically isolate the Majorana
stripe from its surrounding, turning it into an island with a capacitance $C$.
The related charging energy $E_C = e^2 / 2 C$ sets the scale for the maximally
achievable MBS coupling energies. To estimate $E_C$, we assume that the
capacitive coupling of the superconducting layer to the top depletion gate
gives the dominant contribution to the total capacitance $C$. We thus obtain
$C \approx \varepsilon_0 \varepsilon_r W_S L_S / d$ for a parallel-plate
capacitor, where $\varepsilon_0$ is the vacuum permittivity and
$\varepsilon_r$ is the permittivity. Furthermore, $d$ is the thickness of the
aluminum oxide layer separating the superconductor and the top gate, $W_S$
is the width, and $L_S$ the length of the Majorana stripe. Using $\varepsilon_r =
9.1$, $d = 20~\text{nm}$, $L_S = 2~\mu \text{m}$, and $W_S = 200~\text{nm}$
[as in \Fig{fig:s-stripe-control}], we obtain $E_C \approx 25~\mu\text{eV}$.

To reduce the MBS coupling energy, one would couple the superconducting top
layer to a bulk superconductor by a gated semiconductor junction. Based on Eq.~(5) in {\Cite{Aasen16}}, one could expect a coupling energy of
$E_M < 1~$neV for $E_J > 0.75~$meV, while the gap to the excited states is
always larger than $E_C$. With this estimate, we see that charging effects may
provide an alternative to control the coupling of MBSs not only in nanowire
setups, but also in 2DEG structures.

\subsection{Time scales constraints from trijunction
tuning}\label{sec:timescales}

We finally turn to the discussion of the times scales for braiding. A rough
estimate of the time scale $T$ needed to carry out the steps of the braiding
protocol is given by
\begin{eqnarray}
  T_L : = \frac{1}{\min ( E_M, E_g)}  \ll & T & \ll 
  T_U : = \frac{1}{\text{max}  ( E_0)} . 
\end{eqnarray}
Here, $E_0$ is the energy of the MBSs close to zero energy, $E_M$ is the
energy of the fermionic mode formed from coupled MBSs (as sketched in
\Fig{fig:braiding}, at least two MBSs always remain coupled), and $E_g$ is the
energy gap to other excited states.

As we have shown above for both N- and S-stripe designs, the minimal MBS
coupling energies in the trijunction region can be suppressed to $E_0 \lesssim
1~\text{neV}$, which translates into an upper time scale of about
\begin{eqnarray}
  T_U & \gtrsim & 4~\mu \text{s} . 
\end{eqnarray}
Our findings also indicate that coupling energies of $E_M \gtrsim 10~\mu
\text{eV}$ are possible, which are always lower than $E_g$. This yields a
lower time scale of
\begin{eqnarray*}
  T_L & \lesssim & 0.5 \tmop{ns} .
\end{eqnarray*}
This means, as mentioned initially, that the steps of the braiding protocol
can be carried out on a time scale of about 10-100 ns. The coupling energies
of about 10 $\mu$eV correspond to a temperature of about 120 mK. To initialize
the system, one has to cool below this temperature, which is experimentally
demanding but possible. The time-scale window derived here might be even wider
by optimizing the tuning-fork control further, especially with somewhat
increased coupling energies.

\subsection{Conclusion}

In this work, we have shown that the coupling of Majorana bound states (MBSs)
in simple network components formed in two-dimensional electron gases (2DEGs)
can be controlled electrically. MBSs can be created at the ends of topological
quasi-1D channels in two different designs, namely (i) in S-stripes (top Al
stripe above the channel) and (ii) N-stripes (top Al next to the channel).

We showed that the coupling between parallel Majorana stripes can be
suppressed for separations of a few hundred nanometers using potential
barriers on the meV scale. MBSs can further be coupled in trijunction in a
tuning-fork shape. We showed that gate control can be used to tune the
coupling of MBSs within a Majorana stripe and between different stripes in a
range from 1 neV to about 10 $\mu$eV. At the same time, a sizable topological
energy gap can be retained (tens of $\mu$eV), which depends on the potential confining the topological channels.

To realize a trijunction in a 2DEG-based platform, it is important to use a tuning-fork as depicted in \Fig{fig:devices-simulated}(c) and (d) because the topological gap is very sensitive to a misalignment of the stripe and magnetic field direction. It is also important to separate the different stripes from each other through potential barriers of a few meV, i.e., barriers larger than typical induced superconducting gaps. This is needed to avoid uncontrolled couplings between the MBS. The precise dimensions of the tuning fork may, by contrast, vary somewhat and are not tied to the specific example discussed in this paper. However, using sharp potential barriers, the stripe width has to be between 150 - 200 nm to achieve a control over the MBS coupling within a stripe (at zero chemical potential and experimentally motivated parameters used here).

Based on our numerical calculations,
we estimate that an adiabatic braiding protocol can be carried out in a time
window of 10-100 ns. Moreover, the findings of our study are
useful for Majorana box qubits {\cite{Plugge17}} and related network
designs {\cite{Vijay16,Karzig16}} aiming at readout-based braiding. These
proposals involve quantum dots, which could be easily integrated into a
2DEG-based platform. 2DEGs may thus provide an alternative route to build more
complex Majorana devices with the additional advantage of flexibility in
fabrication as compared to other platforms.

\begin{acknowledgments}
{We acknowledge stimulating discussions with M. Kjaergaard, P. Kotetes, C. M. Marcus, F. Nichele, A. Stern, and H. J. Suominen, and support from the Crafoord Foundation (M. L. and M. H.), the Swedish Research Council (M. L.), and The Danish National Research Foundation.}
\end{acknowledgments}

\appendix

\section{Green's function approach}\label{app:gf}

\subsection{Two-dimensional tight-binding model}\label{app:tb}

Our tight-binding model is obtained by discretizing the
continuous spatial coordinates $(x, y)$ as a lattice with points $( x_n, y_m)
= ( n \cdot d, m \cdot d)$, where $d$ denotes the lattice constant.
Due to the discretization, we have to replace the
derivatives by finite differences \cite{Ferry97Book}:
\begin{eqnarray}
  \frac{\partial \psi}{\partial x} (x_n) & \approx & \frac{\psi (x_{n + 1}) -
  \psi (x_{n - 1})}{2 d}, \\
  \frac{\partial^2 \psi}{\partial x^2} (x_n) & \approx & \frac{\psi (x_{n +
  1}) + \psi (x_{n - 1}) - 2 \psi (x_n)}{d^2} . 
\end{eqnarray}
Analogous formulas apply for the $y$ derivative. Using this procedure, we can
rewrite the Hamiltonian~\eq{eq:ham} and the retarded self energy~\eq{eq:sigma} in tight-binding
approximation. Defining the hopping amplitudes
\begin{eqnarray}
  t  \text{  =  } \frac{1}{2 m^{\ast} d^2}, &  & t_{\tmop{SOC}} =
  \frac{\alpha}{2 d},
\end{eqnarray}
they read
\wideeq{
  \begin{eqnarray}
  H & = & \sum_{n_x n_y \tau \sigma}  \{[\tau (4 t - \mu)] |n_x n_y
  \sigma \tau \rangle \langle n_x n_y \sigma \tau | \nobracket + i \sigma E_Z
  / 2| n_x n_y \tau \bar{\sigma} \rangle \langle n_x n_y \tau \sigma |
  \nonumber\\
  &  & \text{\ \ \ \ \ \ }- [\tau t |n^+_x n_y \tau \sigma \rangle 
    \langle n_x n_y \tau \sigma | + \text{H.c.}] - [\tau \sigma t_{\tmop{SOC}}
    |n_x^+ n_y \tau \bar{\sigma} \rangle  \langle n_x n_y \tau \sigma | +
  \text{H.c.}] \nonumber\\
  &  & \text{\ \ \ \ \ \ } - [\tau t \nobracket |n_x n^+_y \tau \sigma
  \rangle  \langle n_x n_y \tau \sigma | + \text{H.c.}] - [i \tau
  t_{\tmop{SOC}} |n_x n^+_y \tau \bar{\sigma} \rangle  \langle n_x n_y \tau
  \sigma | + \text{H.c.}]\} ,  \label{eq:htb}\\
  \Sigma_s^R ( E) & = & \sum_{n_x n_y \tau \sigma}  (Z^{- 1} (n_x, E) - 1) \{ -
  \omega |n_x n_y \sigma \tau \rangle \langle n_x n_y \sigma \tau | \nobracket
  + \Delta |n_x n_y  \bar{\tau} \sigma \rangle \langle n_x n_y \tau \sigma |
  \}  \label{eq:sigmatb}
  \end{eqnarray}
}
Here, $|n_x n_y \tau \sigma \rangle$ denotes a state with an electron ($\tau =
+$) or hole ($\tau = -$) with spin $\sigma = \pm$ localized at lattice point
$(n_x, n_y)$. We introduce the short-hand notations $\bar{\sigma} = - \sigma$,
$\bar{\tau} = - \tau$, $n_i^{\pm} = n_i \pm 1$, and the $Z$ factor
{\cite{SauProximityEffect}}
\begin{eqnarray}
  Z^{- 1} (n_x, E) & = & 1 + \frac{p ( E) \Gamma (n_x)}{\sqrt{\Delta^2 - (E +
  i 0_+)^2}}, 
\end{eqnarray}
with the prefactor
\begin{eqnarray}
  p ( E) & = & \left\{ \begin{array}{ll}
    1 & | E | < \Delta,\\
    i \tmop{sgn} ( E), & | E | > \Delta .
  \end{array}  \right.
\end{eqnarray}
In the limit $E \rightarrow 0$, we obtain $Z^{- 1} \rightarrow 1 + \Gamma /
\Delta$ and the pairing term in Eq.~\eq{eq:sigmatb} is $\propto \Gamma \tau_x$.
However, in nearly all our calculations we keep finite-frequency corrections,
which leads to modifications of the pairing term at finite energy.

\subsection{Self energy of semi-infinite outer
regions}\label{app:surfacegreen}

In our numerical approach, we divide the system into a central part of
interest (lattice points with $| x_n | \leqslant N_x / 2$ and $0 < n_y
\leqslant N_y$) and an outer region that is integrated out. In
this Appendix, we explain how to compute the retarded self energy of the outer
regions, entering into Eq.~(\ref{eq:grc}), by solving a rather simple
eigenvalue problem {\cite{Wimmerphd}}.

Based on Eqs. (\ref{eq:h}) and (\ref{eq:sr}), one can derive a Dyson equation
with $H_{c o}$ as perturbation and one can show that the retarded self energy
can be expressed as:
\begin{eqnarray}
  \Sigma^R_o ( \omega) & = & H_{c o} G_{0, o}^R ( \omega) H_{c o} . 
  \label{eq:sro}
\end{eqnarray}
This expression contains the Green's function of the outer region in the
absence of a coupling to the central region:
\begin{eqnarray}
  G_{0, o}^R ( E) & = & \frac{1}{E - H_o - \Sigma^R_{s, o} ( E)  + i 0_+} . 
  \label{eq:gr0o}
\end{eqnarray}
Since $H_{c o}$ couples only nearest-neighbor sites, one actually only needs
the Green's function $G_{0, o}^R$ evaluated at the sites adjacent to the
central region, $g^R_{0, o} = G_{0, o}^R ( n_y = N_y + 1)$. We assumed her for
simplicity that the outer region covers the lattice sites $n_y > N_y$ as, for
example, in \Fig{fig:devices-simple}.

To derive an expression for $g_{0, o}^R$, we can follow the steps given in
{\Cite{Wimmerphd}} while additionally accounting for the $\Sigma^R_{s,
o}$ by defining a 'Hamiltonian' $\tilde{H}_o = H_o + \Sigma^R_{s, o}$. The
basic idea to compute $g_{0, o}^R$ is to express it terms of the eigenbasis of
$\tilde{H}_o$. For this, one uses that $\tilde{H}_o$ has a tridiagonal form in
the lattice points along the $y$ direction:
\begin{eqnarray}
  \tilde{H}_o & = & \sum_{n_y > N_y} \tilde{H}_{o, 0} | n_y \rangle \langle
  n_y |  \label{eq:htildeo}\\
  &  & \text{ \ \ \ \ \ \ } + H_{o, +} | n_y + 1 \rangle \langle n_y | +
  H_{o, -} | n_y \rangle \langle n_y + 1 | . \nonumber
\end{eqnarray}
Here, $\tilde{H}_{o, 0 }$ and $H_{o, \pm}$ are \ $4 N_x \times 4 N_x$ matrices. For
simplicity, we consider here the semi-infinite region for the parallel stripes
in Figs.~\ref{fig:devices-simulated}(a) and (b); for the tuning-fork setups
$\tilde{H}_o$ has to extend also for $n_y \leqslant 0$. \ Note that
$\Sigma^R_{s, o}$ only modifies the diagonal block $\tilde{H}_{o, 0}$ in Eq.
(\ref{eq:htildeo}). It can then be shown that the surface Green's function
satisfies the relation
\begin{eqnarray}
  g_{0, o}^R & = & U_< \Lambda_< U_<^{- 1} H_{o, -}^{- 1},  \label{eq:gr0o}
\end{eqnarray}
provided $H_{o, -}$ can be inverted, which is possible in our case. The
matrices $U_<$ and $\Lambda_<$ are obtained from the solutions of the
following eigenvalue problem:
\begin{eqnarray}
  \left(\begin{array}{cc}
    0 & 1\\
    - H_{o, -} & E - \tilde{H}_{o, 0}
  \end{array}\right) \left(\begin{array}{c}
    \tmmathbf{u}_n\\
    \lambda_n \tmmathbf{u}_n
  \end{array}\right) & = & \lambda_n \left(\begin{array}{c}
    \tmmathbf{u}_n\\
    \lambda_n \tmmathbf{u}_n
  \end{array}\right) . \nonumber\\
  &  &  \label{eq:genev}
\end{eqnarray}
The eigenvalues in this equation, $\lambda_n = e^{i k_n}$, are connected to
the 'wave vectors' $k_n$ of the eigensolutions of $\tilde{H}_o$ when extended
infinitely in both directions [by omitting the restriction $n_y > N_y$ in Eq.
(\ref{eq:htildeo})]. Of all eigensolutions, one selects those that decay ($|
\lambda_n | < 1$) or propagate ($| \lambda_n | < 1$, $k_n > 0$) in the positive
$y$ direction, which we denote by a subscript '$<$'. The corresponding
eigenvalues fill the diagonal entries of $\Lambda_< = \tmop{diag} (
\lambda_{1, <}, \ldots, \lambda_{4 N_x, <})$ and the eigenvectors are
contained in $U_< = ( \tmmathbf{u}_{1, <}, \ldots, \tmmathbf{u}_{4 N_x, <})$.
There exist exactly $4 N_x$ solutions of type '$<$' because for every solution
$\lambda_n$ of Eq.~(\ref{eq:genev}), also $1 / \lambda_n$ is a solution (not
shown here). Once $g_{0, o}^R$ has been determined from Eq.~(\ref{eq:gr0o}),
it can be inserted into Eq.~(\ref{eq:sro}) and the desired self energy can be
computed.

\subsection{Recursive Green's function method}\label{app:recursive}

In principle, one can compute the Green's function of the central region by
computing the inverse in its definition:
\begin{eqnarray}
  G^R_c ( E) & = & \frac{1}{\omega + i 0 - H_c - \Sigma^R_{s, c} ( E) -
  \Sigma^R_o ( E)} .  \label{eq:grc-app}
\end{eqnarray}
However, computing this inverse is computationally demanding and, in fact, one
does not need the diagonal entries $\langle n_x n_y | G_c | n_x n_y \rangle$
to compute the local density of states, Eq.~(\ref{eq:locdos}). We suppress
here the energy argument of the Green's function to simplify the expressions.
In this Appendix, we show how to compute $\langle n_y | G_c | n_y \rangle$
recursively, where $\langle n_y | \ldots | n_y \rangle$ has to be understood
as taking partial matrix elements.

To start with, we consider the system as a quasi-1D chain in $y$ direction
with each lattice point $n_y$ associated with a $4 N_x$-dimensional subspace.
The idea is to exploit the fact that the inverse Green's function $G_c^{R,-1}$
has block-tridiagonal structure along the chain in $y$-direction:
\begin{eqnarray}
  G_c^{R,-1} & = & \sum_{1 \leqslant n_y \leqslant N_y} A_{n_y} | n_y
  \rangle \langle n_y |  \label{eq:grc}\\
  &  & + B_{n_y}^{\dag} | n_y + 1 \rangle \langle n_y | + B_{n_y} | n_y
  \rangle \langle n_y + 1 | . \nonumber
\end{eqnarray}
We next define
\begin{eqnarray}
  | X^{( n)} \rangle & = & G_c^R | n_y = n \rangle \text{ = } \sum_m X^{( n)}_m | n_y =
  m \rangle, 
\end{eqnarray}
which satisfies by definition the following equation:
\begin{eqnarray}
  G^{R,-1}_c | X^{( n)} \rangle & = & | n_y = n \rangle .  \label{eq:xsolve}
\end{eqnarray}
Our goal is to compute $X^{( n)}_n = \langle n_y = n | G^R_c | n_y = n \rangle$.
For this, we insert Eq.~(\ref{eq:grc}) into Eq.~(\ref{eq:xsolve}) and project
on $\langle n_y = m |$. This yields for $X^{( n)}_m = \langle n_y = m | G^R_c |
n_y = n \rangle$
\begin{eqnarray}
  &  & B_{m - 1}^{\dag} X^{( n)}_{m - 1} + A_m X^{( n)}_m + B_m X^{( n)}_{m +
  1} \nonumber\\
  & = & \delta_{n, m} \mathbbm{1}_{4 N_x \times 4 N_x},  \label{eq:trilinear}
\end{eqnarray}
with the definitions $B_0^{\dag} = B_{N_y} = 0$. One can solve this system of
linear equations by Gaussian elimination. Let us simplicity assume $n \neq 1,
N_y$. Starting from Eq.~(\ref{eq:trilinear}) for $m = 1$, we can first solve
for $X^{( n)}_1$, insert the resulting expression into Eq.
(\ref{eq:trilinear}) for $m = 2$, solve for $X^{ ( n)}_2$, and so on.
Repeating this procedure, one can show the following recursion relation by
induction:
\begin{eqnarray}
  X^{( n)}_m & = & - M^-_m B_m X^{( n)}_{m + 1}, \text{ \ \ \ \ } ( m < n) \\
  M^-_m & = & \frac{1}{A_m - B_{m - 1} M^-_{m - 1} B_{m - 1}^{\dag}}, 
  \label{eq:mminusdef}\\
  M^-_1 & = & \frac{1}{A_1} .  \label{eq:m1def}
\end{eqnarray}
Proceeding analogously when starting from Eq.~(\ref{eq:trilinear}) for $m =
N_y$, one obtains a second recursion relation:
\begin{eqnarray}
  X_m^{( n)} & = & - M_m^+ B_{m - 1}^{\dag} X^{( n)}_{m - 1}, \text{ \ \ \ } (
  m > n) \\
  M_m^+ & = & \frac{1}{A_m - B_m M^+_{m + 1} B^{\dag}_m}, 
  \label{eq:mplusdef}\\
  M^+_{N_y} & = & \frac{1}{A_{N_y}} .  \label{eq:mnydef}
\end{eqnarray}
Finally, using Eq.~(\ref{eq:trilinear}) for $m = n$, we obtain
\begin{eqnarray}
  X^{( n)}_n & = & \langle n_y = n | G_c^R | n_y = n \rangle  \label{eq:xnn}\\
  & = & \frac{1}{- B_{n - 1}^{\dag} M^-_{n - 1} B_{n - 1} + A_n - B_n M^+_{n
  + 1} B_n^{\dag}} . \nonumber
\end{eqnarray}
Note that all the above equations also hold for $n = 1, N_y$. Thus, we have to
iterate Eqs. (\ref{eq:mminusdef}) and (\ref{eq:mplusdef}) starting from Eqs.
(\ref{eq:m1def}) and (\ref{eq:mnydef}) until we obtain $M^-_{n - 1}$ and
$M^+_{n + 1}$, respectively, and insert these into Eq.~(\ref{eq:xnn}) to
obtain the the component we are interested in.

This is an efficient approach to compute the diagonal matrix elements of the
Green's function: While Eq.~(\ref{eq:grc-app}) requires the inversion of a $4
N_x N_y \times 4 N_x N_y$ matrix, the recursive method requires $2 N_y$
inversions of $4 N_x \times 4 N_x$ matrices. Note that the iterations given by
Eqs. (\ref{eq:mminusdef}) and (\ref{eq:mplusdef}) do not have to be repeated
all the way to obtain $X^{( n)}_n$ for different $n$. For example, if $M_{n -
1}^-$ and $M^+_{n - 1}$ are known from calculating $X^{( n)}_n$, then one only
needs to compute $M_n^-$ from applying Eq.~(\ref{eq:mminusdef}) once. The
component $X^{( n + 1)}_{n + 1}$ is obtained using $M_n^-$ and $M^+_{n + 2}$,
where \ $M^+_{n + 2}$ is already known from computing $M^+_{n + 1}$.

\section{Topological phase-transition point for
S-stripes}\label{app:phasetransition}

We derive in \App{app:condition} an approximate formula
for the condition for a topological phase transition in S-stripes. Based in
this, we show in \App{app:voltage-control} that a gate-controlled
phase transition can be evoked for rather small changes in the confinement
potential. This complements the example discussed in the main part
[\Fig{fig:s-stripe-control}], where we focused on an example where rather large
gating potentials are needed.

\subsection{Condition for phase transition }\label{app:condition}

We start from the full Hamiltonian (\ref{eq:ham}) extended to infinity in all
directions. In this case, we can replace $- i \partial_y \rightarrow k_y$,
since the momentum in the stripe direction is conserved. A topological phase
transition point requires a closing of the superconducting gap at $k_y = 0$
{\cite{Hell17a}}. Using the notation of \Sec{sec:model}, the Hamiltonian
reads for $k_y = 0$:
\begin{eqnarray}
  H & = & - \left( \frac{\partial_x^2}{2 m^{\ast}} + \mu ( x) \right) \tau_z +
  i \alpha \sigma_y \partial_x \tau_z + \frac{E_Z}{2} \sigma_y \nonumber\\
  &  & + \Gamma ( x) \tau_x,  \label{eq:ham1d}
\end{eqnarray}
where the effect of the gating is absorbed a position-dependent
electro-chemical potential $\mu ( x) = \mu - \phi ( x)$ with electrostatic potential $\phi ( x)$. The second line
incorporates the superconducting proximity effect through an additional
pairing term, which is the zero-energy limit of the self energy, i.e.,
$\Sigma_R^s ( x, E = 0) = \Gamma ( x) \tau_x$. This replacement introduces no
additional approximation as we are only interested in the position of
zero-energy eigenstates.

To find the eigenenergies of Eq.~(\ref{eq:ham1d}), we next apply a unitary
transformation $U = e^{- i m^{\ast} \alpha \sigma_y}$, which turns the
spin-orbit coupling term into a shift of the electrochemical potential
{\cite{Pientka16}}:
\begin{eqnarray}
  H' & = & U H U^{\dag}  \label{eq:ham1dwithoutsoc}\\
  & = & - \left( \frac{\partial_x^2}{2 m^{\ast}} + \mu ( x) + \frac{m
  \alpha^2}{2} \right) \tau_z + \frac{E_Z}{2} \sigma_y + \Gamma ( x) \tau_x.
  \nonumber
\end{eqnarray}
To obtain a simple estimate of the eigenenergies, we assume that the wave
function is confined by a hard wall in a range $| x | \leqslant \tilde{W}_S$.
We allow this effective width $\tilde{W}_S$ to larger than the width $W_S$ of
the superconducting stripe to account for a nonzero decay length $\lambda$
into the barrier region. A simple estimate is $\tilde{W}_S = W_S + 2 \lambda$.
To simplify the procedure further, we assume that the superconducting
pairing is uniform, $\Gamma ( x) = \Gamma$. This allows us to find the
eigenstates of Eq.~(\ref{eq:ham1dwithoutsoc}) from the ansatz $\tmmathbf{\psi}
( x) = e^{i k_x x} \tmmathbf{\chi}$ with a spinor $\tmmathbf{\chi}$. Solutions
satisfy $k_x = k_{x, n} = n \pi / \tilde{W}_S$, $n \geqslant 1$ with
eigenenergies
\begin{eqnarray}
  E_n & \approx & \pm \frac{E_Z}{2} \pm \sqrt{\Gamma^2 + \left( \frac{k_{x,
  n}^2}{2 m^{\ast}} - \mu - \frac{m^{\ast} \alpha^2}{2} \right)^2} . 
\end{eqnarray}
The lowest bound state for $n = 1$ crosses zero energy under the condition
\begin{eqnarray}
  E_Z & \approx & 2 \sqrt{\Gamma^2 + \left( \frac{\pi^2}{2 m^{\ast} 
  \tilde{W}^2_S} - \mu - \frac{m^{\ast} \alpha^2}{2} \right)^2}.
  \label{eq:ezstar}
\end{eqnarray}
This formula holds strictly only under the condition $\lambda \ll W_S$ and has
been used in the discussion in \Sec{sec:s-stripe}. We note that the
actual superconducting proximity effect should also be weaker than what we
used in the above estimate because the wave function also leaks into the
nonproximitized region. However, the probability density in the
nonproximitized region is rather small, so one would expect the effect of the
effective stripe width to affect mostly the reduced confinement energy.

\FigureSmaller{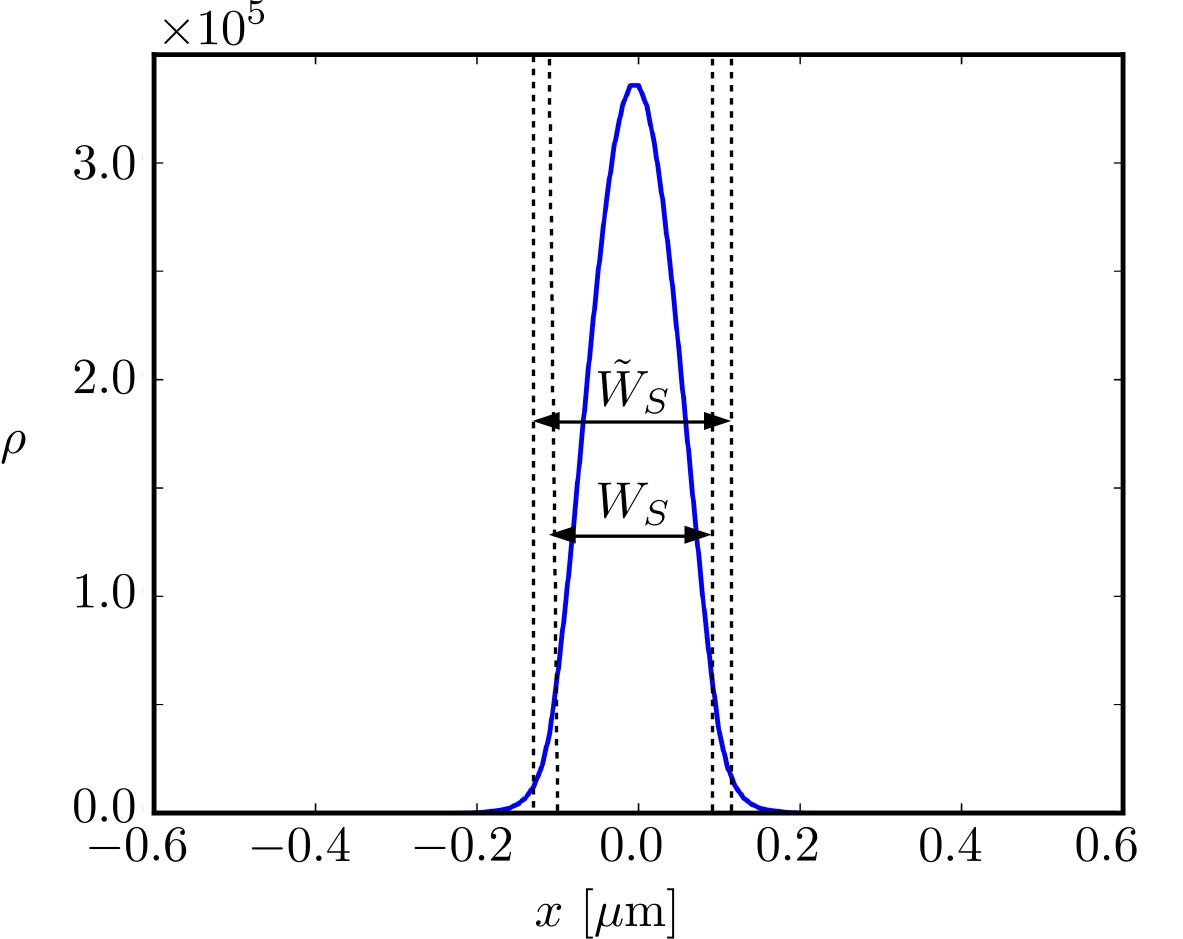}{Effective
  vs.~actual stripe width. The plot is a cut through the lower part of
  \Fig{fig:s-stripe}(c) at $y = 100~\text{nm}$, here shown for the entire
  $x$-range used in the calculation. The actual stripe width $W_S = 200~\text{
  nm}$ and the effective stripe width $\tilde{W}_S = W_S + 2 \lambda \approx
  W_S + \sqrt{2 / m^{\ast} \phi_D} \approx 266~\text{nm}$ are
  indicated.\label{fig:s-stripe-cut}}

\subsection{Gate tunability of phase transition point
}\label{app:voltage-control}

In this Appendix, we briefly show that a gate-induced phase transition can
also occur for smaller potential barrier differences. For this purpose,
we consider in \Fig{fig:s-stripe-control-app} a slightly reduced stripe
width of $W_S = 160~\text{nm}$ as compared to $W_S = 200~\text{nm}$ in
\Fig{fig:s-stripe-control} in the main part. One can clearly see that the
splitting of the Majorana mode already appears when raising the potential barrier to a few
meV [\Fig{fig:s-stripe-control-app}(a) and (b)]. One could use this
insight to optimize the operation of a tuning-fork device: While a large
barrier can be used to separate the two parallel stripes on one side, a rather
soft barrier can be used on the other side to tune the MBS coupling energy
within a stripe. The stripe width then has to be adjusted accordingly.

\Figure{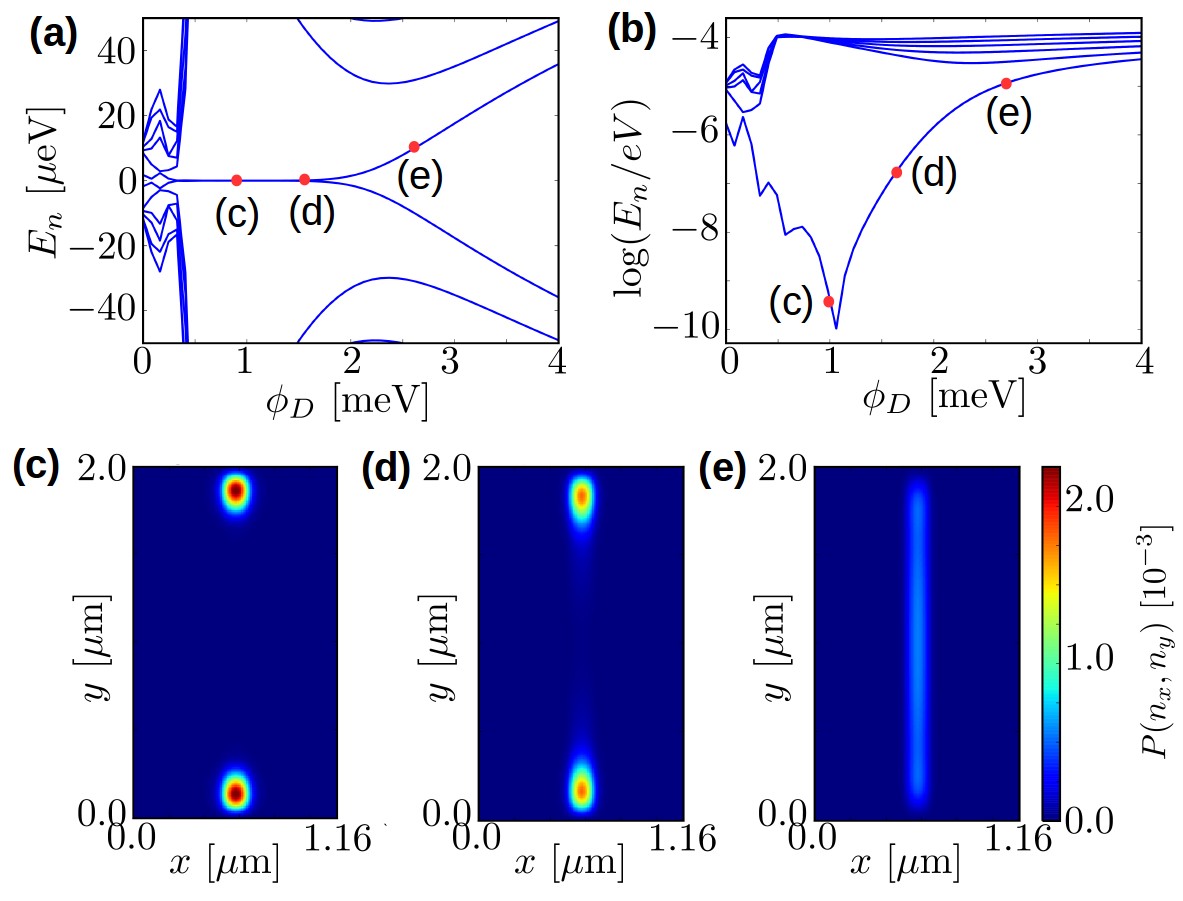}{Gate-tunability
  of the MBS wave function overlap in finite S-stripes. Panels (a) and (b)
  show the resulting energy spectrum as a function of the barrier potential
  $\phi_D$ adjacent to the S-stripe [see \Fig{fig:s-stripe}(a)] on a
  linear and a logarithmic scale, respectively. In (c)--(d), we show the
  probability density $P ( n_x, n_y)$ for the energy eigenstates closest to
  zero energy for different barrier heights $\phi_D$ as indicated in (a) and
  (b). Except for $W_S = 160~\text{nm}$, all parameters are as in
  \Fig{fig:s-stripe-control}.\label{fig:s-stripe-control-app}}

\section{Influence of electrostatic potential profile on the energy
spectrum}\label{app:potential}

In the main part of the paper, we used for all our calculations a rectangular
profile for the electrostatic potential barrier induced by the gates. In this
Appendix, we show that this is a reasonable approximation by computing the
energy spectrum of a single S-stripe [\Fig{fig:potential}(a)] using a
more realistic potential profile. Such Schr{\"o}dinger-Poisson calculations
have already been undertaken for nanowire setups {\cite{Vuik16,Kammhuber17}}
and revealed relevant effects due to a realistic modeling of the electrostatics.
In our case, by contrast, we expect only slight modifications to the energy
spectrum for the parameter regime considered in this paper. We assume,
however, that the electrochemical potential can be tuned freely in the 2DEG
into a low-density regime where MBSs appear.

\Bigfigure{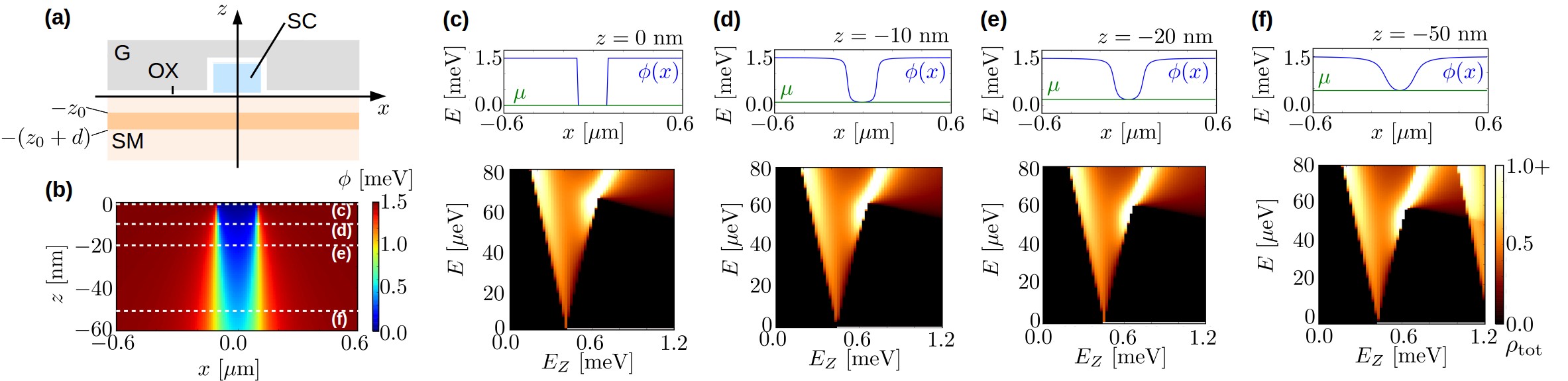}{Influence
  of electrostatic potential profile on energy spectrum. (a) Sketch of the
  cross section of an S-stripe device, consisting of a top gate (G), separated
  from the superconducting stripe (SC) and the semiconductor heterostructure
  (SM) by an oxide layer (OX) whose width is neglected. (b) Electrostatic
  potential $\phi ( x, z)$, Eq.~(\ref{eq:phigate}), induced by a split-gate at
  potential $\phi_G = 1.5~\tmop{meV}$ and a grounded superconducting
  stripe. (c) -- (f), top panels: Horizontal cuts (blue) through the potential
  landscape in (b) along the dashed lines at different distances $- z$ to the
  gate. In our calculations, the electrochemical potential (green) $\mu = \phi
  ( x = 0, - z)$ is adjusted to the bottom of the potential well created by
  the gates. (c) -- (f), bottom panels: total density of states
  $\rho_{\text{tot}}$ as a function of energy and Zeeman energy. Except for
  the modified potential profile and adjusted value of $\mu$, all parameters
  are the same as in \Fig{fig:s-stripe}(d).\label{fig:potential}}

A more realistic modeling of the electrostatic potential is obtained by
solving both the Schr{\"o}dinger equation,
\begin{eqnarray}
  H [ \phi] \psi_n & = & E_n \psi_n,  \label{eq:schroedinger}
\end{eqnarray}
and the Poisson equation,
\begin{eqnarray}
  \varepsilon_0  \vec{\nabla} \cdot \varepsilon \vec{\nabla} \phi & = & - \rho
  [ \{ | \psi_n |^2, \mu - E_n \}],  \label{eq:poisson}
\end{eqnarray}
self-consistently. In Eq.~(\ref{eq:schroedinger}), $H$ is the Hamiltonian
describing in principle the entire device including the 2DEG heterostructure
and the superconductor [\Fig{fig:potential}(a)]. The Hamiltonian
incorporates the effect of the electrostatic potential $\phi ( \tmmathbf{x})$,
which acts as a position-dependent shift of the electrochemical potential $\mu
\rightarrow \mu - \phi ( \tmmathbf{x})$. The electrostatic potential is the
solution of Eq.~(\ref{eq:poisson}), in which $\varepsilon_0$ is the vacuum
permittivity, $\varepsilon$ is the dielectric constant of the material (which
can position-dependent) and $\rho$ is the charge distribution. The charge
distribution, in turn, depends on the probability density associated with the
electronic energy levels filled up to $\mu$ at zero temperature. Solving Eqs.
(\ref{eq:schroedinger}) and (\ref{eq:poisson}) incorporates charging effects
at the Hartree level.

In order to obtain MBSs with a large topological energy gap, the chemical
potential is close to the bottom of the potential well {\cite{Hell17a}}. Under
this condition, the electron charge density in the 2DEG is rather small. In
this case, one can ignore the effect of the electron density on the electrostatic
potential as we argue next. Concretely, let us consider the case that the 2DEG
under the superconductor is occupied only by a single mode due to the
confinement in the $z$ and the $x$ direction [\Fig{fig:potential}(a)]. Since the
stripe length in the $y$ direction is much larger than all other length scales,
the level spacing of modes in the $y$ direction is rather small. The electron
density of states per length in the $y$ direction can thus approximated by that of
infinite 1D system at zero temperature:
\begin{eqnarray}
  n_{\text{1D}} & = & \frac{\sqrt{2 m^{\ast}}}{\pi}  \sqrt{\mu - E_0}. 
\end{eqnarray}
Neglecting superconductivity, we can set $E_0 = - E_{\tmop{SO}} = -118.5~\mu$eV,
the spin-orbit energy, and $\mu = 0$ as in most of calculations. Using $m^{\ast} = 0.023 m$, we obtain
\begin{eqnarray}
  n_{\text{1D}} & \approx & 10^6 / \text{m} . 
\end{eqnarray}
The charge density per volume is then given approximately by $\rho = ( - e)
n_{\text{1D}} / A$, assuming for simplicity that the charge is uniformly
distributed over the cross section area $A = W_S \cdot d_z$ of the 1D channel.
This area is given by the stripe width $W_S \sim 200~\text{nm}$ in the $x$
direction and the width $d_z \sim 10~\text{nm}$ of the quantum well in the $z$
direction. Since $W_S \gg d_z$, \ the electric field $E_{\text{el}}$ will
mostly point along the $z$ direction and only at the edges of the stripe there
will be a contribution in the $x$ direction. We can thus estimate the effect of
the charge density on the potential in the quasi-1D channel by considering the
potential difference $\Delta \phi = \phi ( x = 0, z_0) - \phi ( x = 0, z_0 -
d_z)$ across the quantum well in the $z$ direction [\Fig{fig:potential}(a)].
Using that the electric field points in the $z$ direction, we can integrate Eq.
(\ref{eq:poisson}) and obtain
\begin{eqnarray*}
  \Delta \phi & \approx & - \frac{1}{\varepsilon_0 \varepsilon} 
  \frac{d_z^2}{2} \rho \text{ \ = \ } \frac{e}{2 \varepsilon_0 \varepsilon} 
  \frac{d_z }{W_S} n_{\text{1D}},
\end{eqnarray*}
assuming here that the dielectric constant is uniform. For the InAs / InGaAs /
InAsAs heterostructures used in experiments in {\Cite{Shabani16}}, the dielectric
constants for all materials are about $\varepsilon \approx 15$ and we thus
obtain
\begin{eqnarray}
  \Delta \phi & \approx & 25~\mu\text{eV} . 
\end{eqnarray}
This is indeed negligible compared to gate-induced potentials on the meV
scale. We thus set the charge density $\rho = 0$ in the following.

When the charge density in the 2DEG is neglected and when a uniform
dielectric constant is assumed, the gate-induced potential can be computed
analytically for a split-gate structure as shown in
\Fig{fig:potential}(a). We will further make the assumption that the oxide
layer thickness is negligible so that we may can employ Dirichlet boundary
conditions $\phi ( | x | \leqslant W_S / 2, z = 0) = 0$ and $\phi ( | x | >
W_S / 2, z = 0) = \phi_G$ on the $x y$ plane. Using von-Neumann boundary
conditions $\partial \phi ( \tmmathbf{x}) / \partial n = 0$ for $|
\tmmathbf{x} | \rightarrow \infty$, $z < 0$, for the rest of the boundary, the
solution to Eq.~(\ref{eq:poisson}) reads {\cite{Davies95}}:
\begin{eqnarray}
  \phi ( \tmmathbf{x}) & = & \phi_G \left[ 1 + \frac{1}{\pi} \sum_{p = \pm}
  \text{arctan} \left( \frac{W_S + p x}{z} \right) \right].  \label{eq:phigate}
\end{eqnarray}
This gate-induced potential is shown in \Fig{fig:potential}(b) and
horizontal cuts at different distances below the gates are shown in the top panels of
\Fig{fig:potential}(c)--(f). In experimental setups, the 2DEG is about
10-20 nm below the semiconductor surface, for which the shape of the quantum
well is still close to rectangular [Figs.~\ref{fig:potential}(d) and (e)].

We used the more realistic gate potential (\ref{eq:phigate}) for different
fixed values of $z$ to compute the total density of states [bottom panels of
\Fig{fig:potential}(c)--(f)] for otherwise the same parameters as in \Fig{fig:s-stripe}(d). We adjusted the chemical potential in all cases to the
bottom of the potential well, i.e., $\mu = \phi ( x = 0, z)$. For $z = - 10~\text{nm}$ and $z = - 20~\text{nm}$, we can see that there are only minor
differences in the energy spectrum when using a rectangular or a smoothened
potential profile [compare Figs.~\ref{fig:potential}(d) and (e) to
  \Fig{fig:potential}(c)]: The phase-transition point remains nearly unshifted
and the topological energy gap is only slightly reduced.

A qualitative difference in the energy spectrum can only be seen for the
largest depth $z = - 50~\text{nm}$ below the surface [bottom panel of \Fig{fig:potential}(f)]. Here, a second Andreev bound state approaches zero
energy for large Zeeman energies. Such a state would probably also appear for
the other cases of smaller |$z$| when increasing the Zeeman energy. The reason
why this state appears for lower Zeeman energies in the case of $z = - 50~\tmop{nm}$ is probably that the confinement energy for higher-lying excited
states is reduced for the more smoothened potential profile. However, a single
MBS appears still in a broad regime of Zeeman energies. The case of $z = - 50~\tmop{nm}$ is still relevant because it can be used to estimate the effect of
an oxide layer with a nonzero thickness. Our results show that even in this
case, one can expect that the results of the rectangular potential barrier
hold in the single-subband regime with only small modifications.\\

\bibliographystyle{apsrev}
\end{document}